\documentclass[superscriptaddress,column,showpacs,a4paper,
amssymb,amsmath,nobibnotes,aps,prd,longbibliography,
showkeys,
nofootinbib,notitlepage]{revtex4-1}
\usepackage{graphicx}
\usepackage{float}
\usepackage{epsf}
\usepackage{bm}
\usepackage{amsmath}
\usepackage{amsfonts}
\usepackage{amssymb}
\usepackage{epstopdf}
\usepackage{natbib}
\usepackage{color}%
\usepackage{mathrsfs}
\providecommand{\U}[1]{\protect\rule{.1in}{.1in}}

\newcommand{\be}{\begin{equation}}
\newcommand{\ee}{\end{equation}}

\newcommand{\mincir}{\raise
-3.truept\hbox{\rlap{\hbox{$\sim$}}\raise4.truept\hbox{$<$}\ }}
\newcommand{\magcir}{\raise
-3.truept\hbox{\rlap{\hbox{$\sim$}}\raise4.truept\hbox{$>$}\ }}

\newtheorem{remark}{Remark}[section]
\usepackage{xcolor}
\usepackage{hyperref}
\hypersetup{
    colorlinks=true, 
    linkcolor=blue,
    citecolor=magenta}

\begin{document}

\title{
Cartographing Gravity: Operational Maps, the Conatus of Proper Time, and the Potentia of Matter}

\author{Jaume de Haro}
\email{jaime.haro@upc.edu}
\affiliation{Departament de Matem\`atiques, Universitat Polit\`ecnica de Catalunya, Diagonal 647, 08028 Barcelona, Spain}

\author{Emilio Elizalde}
\email{elizalde@ice.csic.es}
\affiliation{Institut de Ciències de l'Espai, ICE/CSIC and IEEC, Campus UAB, C/ Can Magrans, s/n, 08193 Bellaterra, Barcelona, Spain}

\begin{abstract}

Classical natural philosophy sought the natural law governing the motion of free bodies long before daring to construct the appropriate mathematical formalism. Following this longstanding methodological perspective, in the present work 
General Relativity is {reconstructed} from operational and dynamical principles, rather than {taking spacetime geometry as its starting point.}
The \textit{conatus} of a massive free particle is identified with its natural tendency {to follow a worldline of stationary proper time, which, under the appropriate local conditions, corresponds to a maximum of its proper time},
whereas gravitation is interpreted as the \textit{potentia} of matter to modify the local rate of proper time. By
requiring the exact validity of Galilean free fall,  the weak-field static invariant is determined. The gravitational field of moving sources is subsequently obtained by Lorentz-transforming the elementary static solution and causally superposing the retarded contributions of infinitesimal source elements. 
The resulting {linearized field equations coincide with those of} General Relativity in the harmonic gauge.
Finally, imposing general energy-momentum conservation together with the requirement of second-order field equations yields---
{within the class of theories satisfying the stated assumptions and} 
by virtue of Lovelock's theorem---the unique {covariant} completion represented by Einstein's field equations.
Within this {construction, geometry is not postulated from the outset; instead, it emerges as the final mathematical synthesis of operational dynamics based on proper time, inertia, and free fall.}
\end{abstract}

\vspace{0.5cm}

\pacs{01.70.+w, 01.65.+g, 04.20.Cv, 
04.20.-q}

\keywords{Gravitational Red-Shift; Equivalence Principle; Proper Time.}

\maketitle

\section{Introduction}

Our purpose in this essay is to {show how to} reconstruct the weak-field structure of General Relativity from operational dynamical principles rather than from geometry. Our strategy is to chart an alternative route for incorporating gravitation into Special Relativity, namely a route grounded in physical and operational principles rather than in the foundational postulate of spacetime curvature. Historically, the triumph of General Relativity was accompanied by a drastic epistemological shift: the highly abstract, absolute differential calculus of the Göttingen school and the Berlin academy progressively overshadowed the operational, more physically minded foundations of the theory. What began as the fruits of Einstein's deeply physical insights---such as his 1905 mass-energy equivalence and local clock desynchronization---was rapidly absorbed into a highly geometric mathematical framework. We seek here to revisit this historical development by {exploring how} the elegant geometric presentation of General Relativity can be understood not as the only possible starting point, but as a particularly advantageous mathematical construct that emerges naturally from underlying relational dynamics at the final stage of the construction.

Quite probably, this tendency to establish physical theories through axioms, paralleling what is done in mathematics, started in Einstein's case right after his exceptional finding that the already existing Lorentz and Poincaré transformations could be derived from two basic principles: the Principle of Relativity and the constancy of the speed of light in vacuum. They constituted, since that day, the two grounding principles of his Special Theory of Relativity \cite{Eli22}. No wonder then, that he tried to repeat the same conceptual frame when building the general theory, starting, in addition, from the Equivalence Principle. In this latter case, the development of the field equations led him to adopt {the differential-geometrical framework needed to formulate the corresponding gravitational dynamics}. This sort of axiomatic, straightforward approach is certainly operationally sound and very efficient (forgetting for a moment about the enormous difficulty in solving the resulting equations, in many cases). However, following this way, one may lose some of the deep physical insight present in other fundamental theories. This was explicitly recognized by Einstein himself, some years later, on various occasions, as we are going to see in the present paper.

Our alternative construction starts by restoring the primacy of the territory on the map \cite{Korzybski1933}. We initiate the program with the universality of Galilean free fall, embodying the intrinsic nature of inertia. From there, we adopt as a fundamental principle that in the absence of non-gravitational forces, {massive} bodies 
spontaneously evolve along trajectories that {extremize (locally maximize the number of ticks of their own clock)}
 their experienced proper time. This {extremization} represents the physical manifestation of the Spinozian classical \textit{conatus}---the innate endeavor of matter to persevere in its own state---translated here into the relativistic domain. Although Lorentz invariance ensures strict compatibility with relativistic kinematics, the retarded Newtonian potential, as introduced by Max Abraham in 1911 \cite{Abraham1911}, supplies the necessary dynamical content for the gravitational interaction without invoking the accelerated reference frames traditionally required by Einstein's original Equivalence Principle. With these relational elements in place, the invariant interval $ds^2$---strictly understood as the proper time of local clocks \cite{Minkowski1909}---arises naturally for a static gravitational field. By applying kinematic Lorentz transformations to individual mass elements, this formulation extends to moving matter, {and in the weak-field regime the resulting linearized field equations coincide with those of General Relativity in the harmonic gauge}.

The conceptual and epistemological significance of this approach is twofold. First, it treats proper time---rather than abstract manifold curvature---as the primary operational quantity governing the gravitational interaction. As a consequence, in this framework gravitation is no longer taken to be fundamentally geometric; it is simply revealed as the \textit{potentia} of matter to dynamically modulate the internal rhythm of time. Here, geometry does not precede dynamics; it is born from them. Second, as already developed in \cite{Haro2026}, when this linear reasoning is extended beyond the weak-field approximation to ensure general energy-momentum conservation, a deep structural consistency emerges. {The nonlinear completion can then be characterized, within the stated assumptions concerning locality, covariance, and second-order field equations, by the mathematical structure singled out by Lovelock's theorem \cite{Lovelock1971}, leading to Einstein's field equations.} The field acts as its own source, and the resulting non-linearity arises not from an \textit{a priori} geometric axiom, but as {a requirement of dynamical consistency within the specified field-theoretic framework}, as later championed by Feynman, Weinberg, Deser, and others \cite{Feynman1995}.

{

Part of the technical construction presented in this essay builds directly on our previous work \cite{Haro2026}. In particular, the derivation of the proper-time invariant in a static gravitational field, its extension to moving matter, and its agreement with linearized General Relativity in harmonic coordinates were already established there. These results are therefore not claimed as new here, but are incorporated into the broader conceptual framework developed in the present work.

The main developments of the present essay concern the physical and epistemological interpretation of this construction. We introduce the notions of \textit{conatus} and \textit{potentia} as dynamical concepts and relate them to D'Alembert's principle, inertia, gravitation, and the operational meaning of proper time. We also revisit Einstein's rotating-disk argument and examine the distinction between operational geometry and the physical state of the material systems involved. More generally, we clarify how the operational and dynamical principles governing free fall lead to a geometrical description, and under which additional assumptions this transition is justified.

{

The essay does not claim to derive General Relativity without prior assumptions, nor to establish an unrestricted uniqueness theorem. Its aim is instead to exhibit a logically explicit operational route through which the weak-field content of General Relativity can be reconstructed from dynamical and operational principles, with the geometrical description appearing only at a later stage as a compact synthesis of the 
operationally reconstructed relations.
 The use of restricted ansätze is deliberate:
 as illustrated, for instance, by Einstein's early development of General Relativity
  (see, for instance, Einstein's 1912 paper \cite{Einstein1912}), an ansatz is used here as an intermediate representation constrained by physical requirements, rather than as an arbitrary postulate of the final theory. The novelty of the present work is therefore primarily conceptual and foundational. It lies in the proposed ordering and interpretation of the principles—conatus, potentia, proper time, inertia, and gravitation—and in the resulting view of spacetime geometry as an emergent representation of operationally reconstructed dynamical relations, rather than as a foundational premise.

}

}

In summary, it is not our aim to deny the great mathematical elegance of the geometric formalism, but to deontologize it and restore its genuine operational meaning. Following Henri Poincaré's nominalist insight that no geometry can be inherently `true,'' but just `more convenient,'' \cite{Poincare1902} the curved pseudo-Riemannian manifold should be viewed as a compact and very advantageous representation of relational measurements---i.e.\ time intervals, rod contractions, and signal propagations---rather than, literally, as a flexible substance endowed with curvature.

Crucially, this does not imply adopting a radical relationalism that dispenses with background structures altogether: the mathematical manifold remains an indispensable frame required to locate events, represent fields, and define kinematics. However, this formal requirement must not be reified into an autonomous physical entity.

Echoing Newton's \textit{hypotheses non fingo}, we will refrain from presenting metaphysical hypotheses about what gravity ``is'' in itself; we will instead describe its physical effects in terms of the modulation of proper time within the operational construction developed here, whose weak-field reference structure is Minkowskian. This viewpoint is consistent with several remarks made by Einstein himself during his later years. In his 1920 Leiden lecture \cite{Einstein1920}, he warned against treating spacetime as an elastic medium, and in 1952 \cite{Einstein1952} he claimed that spacetime does not have an independent existence, but exists only as a structural quality of the field.

Eventually, General Relativity returns in this way to its conceptual {foundations}: not as a substantive theory about the geometry of a flexible container---where Newton's rigid arena is merely replaced by a deformable spacetime that tells matter how to move while matter tells it how to curve---but as relational dynamics of time and inertia, formulated over a necessary, yet non-reified, mathematical stage.

\

Throughout this work, we adopt units where $ c  = 1$ and $G$ is Newton’s gravitational constant.

\section{D'Alembert and Equivalence Principles}\label{sec-PE}

Newton's second law can be stated as  \cite{Newton1687}:
\begin{quote}
  \textit{The alteration of motion is ever proportional to the motive force impress'd; and is made in the direction of the right line in which that force is impress'd.}
\end{quote}
 In mathematical form, it can be formulated as follows:
\begin{equation}\label{secondlaw}
    {\bf F}dt=d(m{\bf v})\Longleftrightarrow
    {\bf F}d t-d(m{\bf v})=0,
\end{equation}
and, in modern language, as: 
\begin{quote}
    \textit{The balance between the transferred momentum and the variation of linear momentum is always zero,}
\end{quote}
in the same way that we may say: 
"the amount of money deposited into my bank account is equal to the change in my account balance."

\begin{remark}
We should recall that Newton's second law was originally formulated with respect to an  "idealized" background. In principle, its application requires referring the momentum $m{\bf v}$ to Newton's absolute space and absolute time. 
The law may then be reformulated in any other reference frame by introducing the appropriate coordinate transformation (and, where necessary, the corresponding inertial terms).

However, in practice, this methodology faces an obvious difficulty. Our laboratory is at rest or moves uniformly with respect to this idealized frame. But 
the Earth itself rotates and accelerates, so that the law can only be applied approximately with respect to an ideal inertial frame.

From a nominalist perspective, this is not a problem, actually. Echoing Niels Bohr's insight that physics concerns not how nature is, but what we can say about it, physics does not describe reality itself; rather, it constructs maps that allow us to navigate reality. The map is not the territory: it is a simplified representation that enables us to describe and predict phenomena.
Absolute space is therefore understood as part of the Newtonian map rather than as an ontological entity. Within this map, the law is perfectly well defined, even though no laboratory can ever coincide exactly with an inertial frame.
\end{remark}

If we now take Eq. (\ref{secondlaw}), dividing it by $dt$,  we obtain what is known as D'Alembert's Principle \cite{Dalembert1743}:
\begin{equation}
    {\bf F}-\frac{d(m{\bf v})}{dt}=0\Longleftrightarrow {\bf F}-m{\bf a}=0.
\end{equation}
Here, the force ${\bf F}$ is interpreted as the rate at which momentum is transferred to the body, and the \textit{inertial force}, namely $-m{\bf a}$, is the opposite of the rate at which linear momentum changes.

\begin{remark} The momentum balance leads to the 
 Work-Energy Principle
\begin{eqnarray}\label{work-energy}
    {\bf F}\cdot d{\bf r}=d\left(\frac{1}{2}m{\bf v}\cdot {\bf v}\right)\Longrightarrow
    {\bf F}\cdot d{\bf r}={\bf v}\cdot d(m {\bf v}).
\end{eqnarray}
In fact, multiplying equation (\ref{secondlaw}) by ${\bf v}=\frac{d{\bf r}}{dt}$, one finds (\ref{work-energy}).
\end{remark}

For an observer with mass $m_1$ and fixed to the Earth ---where, applied to him, the law yields $N-m_1g=0$, with $N$ being the normal force exerted by the Earth on the observer--- observing an object of mass $m$ falling along the $z$-axis
("taking the $z$-axis pointing vertically upwards"), then the law becomes:
\begin{equation}
    -mg-m\ddot{z}=0.
\end{equation}
In contrast, for a freely falling observer using its own coordinates $\bar{z}=z+\frac{1}{2}gt^2$, the law becomes:
\begin{equation}
    -m\ddot{\bar z}=0.
\end{equation}
Thus, in the freely falling frame, there is neither an inertial force nor gravity.

Let us now consider the rocket thought experiment. Imagine we have a rocket of mass $M$ with no occupants inside, whose engine provides acceleration ${\bf b}$ relative to an inertial observer. The application of this law by the inertial observer yields ${\bf F}_1-M{\bf b}=0$, where ${\bf F}_1$ is the force applied by the engine to the rocket's floor.

If we now place an observer of mass $m_1$ inside the rocket standing on its floor, the law becomes the following
\begin{equation}
    {\bf F}_2-(M+m_1){\bf b}=0,
\end{equation}
where ${\bf F}_2$ is now the force applied by the engine to the rocket's floor to maintain acceleration ${\bf b}$. 

Subtracting the balance equation of the empty rocket from that of the occupied rocket gives:
\begin{equation}
    {\bf N}-m_1{\bf b}=0,
\end{equation}
where ${\bf N}={\bf F}_2-{\bf F}_1$.

If ${\bf i}$ denotes the direction of motion of the rocket, we have ${\bf N}=N{\bf i}$ and ${\bf b}=b{\bf i}$, so the law reads $N-m_1b=0$, which is precisely the same law obtained for an observer standing at rest on the Earth
\begin{eqnarray}
N-m_1g=0.
\end{eqnarray}

Note also the law for an object of mass $m$ initially located at distance $z_0$ from the rocket's floor. As written by the observer standing on the rocket's floor, the law reads $-mb-m\ddot{z}=0$. On the other hand, for the external inertial observer, the object of mass $m$ remains stationary (or moves with uniform linear motion) ---it is the rocket floor that accelerates toward the object--- and therefore the law is expressed as:
\begin{equation}
    -m\ddot{\bar z}=0.
\end{equation}

This illustrates Einstein's Equivalence Principle in its original (1907) form \cite{Einstein1907}: a uniformly accelerated frame in empty space is locally indistinguishable from a frame at rest in a uniform gravitational field.

\begin{remark}
The historical question concerning the original Equivalence Principle---famously rephrased by Feynman \cite{Feynman1964}---asks: \textit{is the rocket accelerating with respect to what?} In retrospect, we could formulate this question in terms of an ``inertial'' reference structure, but not in terms of Minkowski spacetime, since Minkowski's seminal lecture was given in 1908 and his subsequent work was published in 1909 \cite{Minkowski1909}---a geometric formulation that Einstein himself initially dismissed as an unnecessary mathematical abstraction. Einstein need not have interpreted this question as referring to an underlying inertial background. From a coordinate-based, operational perspective, he simply specified the reference system with respect to which the motion was described. The modern question of what the rocket accelerates with respect to may suggest an underlying reference structure, but there is no need to attribute such an ontological background to Einstein's 1907 argument. His construction can be understood entirely in terms of reference systems and the physical equivalence of their local observations.

In Einstein's early relativistic thinking, what was physically primary was the reference system and the operational relations associated with it. In Special Relativity, this physical perspective is embodied by a closed symmetry group—the Poincaré (or Lorentz) group—that maps the class of physical inertial observers onto itself. However, when attempting to extend the principle of relativity to arbitrary non-inertial motion, it proves impossible to isolate a well-defined, restricted group corresponding to ``all physical reference systems.'' The most natural and pragmatic way forward is to extend invariance to the largest possible group: all arbitrary smooth coordinate transformations. It is precisely at this juncture—replacing an unidentifiable subgroup of physical observers with the full diffeomorphism group—that physics naturally transitions into differential geometry, replacing physical reference frames with local coordinate charts on a smooth manifold. Consequently, while covariance was initially understood as a formal requirement on the form of physical laws under coordinate transformations (if a law is expressed as $H_{\mu\nu}=0$ in one system, it retains the form $H'_{\mu\nu}=0$ in another), the modern differential-geometric interpretation goes further, identifying such equations as coordinate-independent (intrinsic) geometric relations whose component expressions in different coordinate systems are merely different representations of the same underlying object.

Coming back to the original Equivalence Principle, this immediately leads to a conceptual tension. From a Newtonian viewpoint, one may introduce an inertial reference structure relative to which an observer at rest in a uniform gravitational field is stationary, whereas a freely falling observer is accelerating with respect to it. Yet the latter is precisely the observer whose motion is inertial in the modern relativistic sense. 

In this sense, General Relativity does not merely extend the Newtonian notion of inertial motion by allowing ``straight'' trajectories to be recovered locally. Rather, it replaces the notion of straight-line motion with the more general dynamical notion of free motion as extremal proper-time evolution, whose mathematical representation is a geodesic. Straight-line inertial motion in Newtonian mechanics and Special Relativity is then recovered as the limiting case in which the gravitational field is absent.

From the present epistemological perspective, 
to describe nature locally, we inevitably construct a
mathematical map containing some reference structure against which
positions, velocities, accelerations, and temporal evolution can be
specified. Such a structure may be indispensable to the description
without being an autonomous physical entity. It is therefore useful to
distinguish the virtual background required by the map from the physical
relations represented by it.

The subsequent development of relativity can then be understood as a
progressive transformation of this reference structure. In Special
Relativity, inertial motion is represented relative to Minkowski spacetime.
In General Relativity, the metric itself becomes dynamical—or, more precisely, as demonstrated by the hole argument \cite{Stachel2014}, an entire equivalence class of metrics represents a single physical situation, so that no particular representative is physically privileged. Nevertheless, a {concrete mathematical description still requires a choice of representative from this equivalence class}. Once a particular representative is chosen by fixing a gauge, 
 {the resulting representation contains} a definite Lorentzian metric with respect to which coordinates, trajectories, and local inertial frames can be described. {This choice does not introduce a non-dynamical background metric: the chosen metric remains a representative of the dynamical gravitational field.} Local inertial frames can then be constructed through Riemann normal coordinates, in which the metric is locally Minkowskian to first order in the coordinates near the chosen event. In this representation, the free motion described above takes its locally Minkowskian form. {The chosen representative is therefore part of the mathematical description, but it does not acquire an absolute physical status; its physical content is relational.
}

For asymptotically flat solutions, however, an additional reference
structure reappears at infinity. For example, in the Schwarzschild
solution, the metric approaches the Minkowski metric asymptotically,
thereby providing an asymptotic inertial structure relative to which the
global state of the isolated source can be defined. This is closely
analogous to the role played by an ideal inertial structure in Newtonian
mechanics: in the description of an isolated gravitating system, one may
operationally approximate such a structure in a region sufficiently far
from the sources, where their gravitational influence becomes negligible.
Thus, even in General Relativity, the statement that a source is ``at
rest'' or that it possesses a given state of motion is meaningful only
relative to some reference structure. Such a structure is indispensable
to the mathematical description, but need not be interpreted as an
autonomous physical background.

In both Newtonian mechanics and General Relativity, an isolated
dynamical system must be situated within a reference structure in order
for its state of motion to be specified. What changes from Newton to
General Relativity is not the need for such a structure in the
mathematical description, but its ontological status.
\end{remark}

To be even more clear, 
let us now combine the Earth and rocket scenarios, considering an elevator on Earth. Suppose that a body of mass $m$ is placed inside an elevator that moves up or down, only. For an observer fixed to the Earth's surface, the balance law reads:
\begin{equation}
    N-mg-m\ddot{z}=0,
\end{equation}
where $N$ is the upwards normal force exerted by the elevator floor on the body. Although the gravitational force acting on the body is always 
$-mg$, the quantity measured by a scale is the normal force N, usually called \textit{apparent weight}.

When the elevator descends with acceleration $\ddot{z}=-a$ ($0<a<g$), the apparent weight of the body registered by a scale on the elevator's floor, 
\begin{eqnarray}
    N=m(g-a),
\end{eqnarray}
decreases. 
In the limiting case $a=g$, corresponding to free fall, the scale reads $N=0$, even though the gravitational force $-mg$
is still present. Conversely, if the elevator accelerates upwards, the apparent weight increases.

If we now view the law from the perspective of a freely falling observer, since gravity disappears in this frame, the law becomes:
\begin{equation}
    N-m\ddot{\bar{z}}=0.
\end{equation}

Since the elevator has acceleration $z=g-a$ relative to the free-falling observer, we recover exactly the same scale,
\begin{eqnarray}
    N=m(g-a).
\end{eqnarray}

Thus, both observers predict the same apparent weight, although they attribute it to a different force balance: the observer on the Earth's surface includes gravity explicitly, whereas the freely falling observer does not.

\begin{remark}
This comparison highlights a profound conceptual distinction, concerning free
fall. In the standard Newtonian interpretation, gravity is treated as an
ontological force, $-mg$, acting continuously on the body, independently of
its state of motion. Consequently, the weightlessness experienced during
free fall is usually regarded as an apparent phenomenon: the gravitational
force remains present, while the supporting force exerted by the surrounding
matter vanishes. Since all parts of a sufficiently small free-falling body
share the same acceleration, no macroscopic contact stresses are required
to maintain its motion.

From an operational and d'Alembertian perspective, however, the quantity
directly accessible to a local mechanical measurement is not the gravitational
term by itself, but the contact force $N$. A scale measures $N$, and the
mechanical stresses experienced by the material body are likewise determined by
contact interactions. Thus, in free fall, to say that
\begin{eqnarray}
N=0,
\end{eqnarray}
is not merely a statement about the absence of an apparent sensation of
weight; it is an objective local physical condition characterizing the body's motion.

In this sense, the scale and the human body provide two physically different,
but conceptually analogous, local detectors of weightlessness: both respond
to the mechanical stresses associated with contact forces. The vanishing of
$N$ therefore has an operational meaning that is independent of the observer's
description of the motion. An observer at rest on the Earth's surface
describes this condition through the balance
\begin{eqnarray}
N-mg-m\ddot{z}=0,
\end{eqnarray}
with $\ddot{z}=-g$.
Whereas a freely falling observer describes the same local condition as
\begin{eqnarray}
N-m\ddot{\bar z}=0,
\end{eqnarray}
with $\ddot{\bar z}=0$ for the free-falling body. The two descriptions
differ in their decomposition into gravitational and inertial contributions,
but agree on the locally measurable contact force.

The physical content of weightlessness is therefore not that gravity has
become ``invisible'' in a merely subjective sense, but that the local
mechanical interaction with the supporting environment has vanished. The
same physical condition, $N=0$, admits two different d'Alembertian descriptions. For the observer at rest on the Earth's surface, weightlessness
is represented by the exact balance between the gravitational and inertial
contributions,
\begin{eqnarray}
-mg-m\ddot z=0,
\qquad \ddot z=-g,
\end{eqnarray}
whereas for the freely falling observer, it is represented by the vanishing
of the inertial contribution, namely
\begin{eqnarray}
-m\ddot{\bar z}=0.
\end{eqnarray}

Thus, the two observers disagree on the decomposition of the balance into
gravitational and inertial terms, but agree on its operational content:
the contact force vanishes, indeed
\begin{eqnarray}
N=0.
\end{eqnarray}

This provides a direct d'Alembertian interpretation of the Equivalence
Principle: different observers may assign different dynamical
decompositions to the same motion, while the locally measurable balance remains unchanged.
\end{remark}

From this perspective, D'Alembert's principle is not merely an alternative formulation of a modern reading of Euler's formulation of Newtonian mechanics; usually written as $\mathbf{F}=m\mathbf{a}$, although Euler's original formulation was $(dt)^2\mathbf{F}=md(d\mathbf{r})$ with $\mathbf{r}=(x,y,z)$ \cite{Euler}. Rather, it replaces a causal relation---in which the applied force is regarded as the cause and the acceleration as its effect---with a balance law in which the inertial term enters on the same footing as the forces applied. The form of this balance is preserved under reference frame 
changes, even though individual terms need not retain the same form, and the transformation is not covariant in the tensorial sense. Different observers may therefore construct different maps of the forces and of the inertial term, while the underlying balance structure does remain unchanged.

\

{
A final remark concerns the mathematical structure underlying Euler's notation. Rather than claiming a historical derivation, we can reconstruct, in a simple geometrical way, the second-difference structure that corresponds to the quantity \(d(dz)\) appearing in Euler's formulation.

From a purely analytical standpoint, the differential relation appearing in Euler's formulation can be recovered in a couple of steps by treating \(dt\) as a constant time increment.}
In the case of  free fall, 
since the velocity is defined as  $v = dz/dt$, one has $dz = v dt$, and taking the differential yields
\begin{eqnarray}
d(dz) = d(v \, dt) = dv \, dt.
\end{eqnarray}

Using now Newton's relation for the change of momentum
$mdv+Pdt=0$, where $P$ is the particle's weight, immediately leads to
\begin{eqnarray} 
m  d(dz) = -P (dt)^2.
\end{eqnarray}

This differential manipulation is algebraically straightforward, but it hides the physical and geometric meaning of the second spatial differential $d(dz)$. 

To make explicit the geometrical structure underlying the analytical formulation of Euler's mechanical principle, let us return to Galileo's geometrical treatment of free fall.
Consider a vertical axis with origin $O$.
Moving downward along this axis, define a sequence of points corresponding to equal discrete-time steps:
$
A_1 = {\Delta t}, \quad A_2 = 2  {\Delta t}, \quad \dots, \quad A_k = k{\Delta t}, \dots
$
At each point $A_k$, erect now a vertical segment of length $gk{\Delta t}$, terminating at $B_k$, for $k=1,\dots,n,\dots$.

Now, consider the right triangle $OA_kB_k$. Its area, denoted by $a_k$, is
\begin{eqnarray}
a_k=\frac{1}{2}A_kB_k
=\frac{1}{2}(k{\Delta t})(gk {\Delta t}  )
=\frac{1}{2}gk^2(  {\Delta t}          )^2,
\end{eqnarray}
{and, for a body freely falling from rest, this area represents the spatial displacement
$z(0)-z( k{\Delta t}   )$.

}

Following Galileo's law of odd numbers, 
let us consider the successive finite spatial increments
{$ \Delta z_k=z((k+1) {\Delta t})-z(k {\Delta t}    ).$}

The difference between two successive increments is, therefore,
{$
{\Delta^2}z_k={ \Delta}z_{k+1}- {\Delta}    z_{k},
$}
or, equivalently,
{\begin{eqnarray}
{\Delta^2}z_k
=z((k+2)   {\Delta t}          )-2z((k+1)  {\Delta t}      )+z(k   {\Delta t}    ).
\end{eqnarray}}

Subtracting now $z(0)$ throughout, this can be expressed directly in terms of the triangle areas, as
\begin{align}
{\Delta^2}z_k
=[z((k+2) {\Delta t}     )-z(0)]
-2[z((k+1)   {\Delta t}       )-z(0)]
+[z(k    {\Delta t}      )-z(0)]
=-(a_{k+2}-2a_{k+1}+a_k).
\end{align}

Thus, geometrically, $- {\Delta^2}z_k      $ is obtained by subtracting twice the area of triangle $OA_{k+1}B_{k+1}$ from the area of triangle $OA_{k+2}B_{k+2}$, and then adding the area of triangle $OA_kB_k$.

Since
$
a_k=\frac12gk^2(    {\Delta t}      )^2,
$
we obtain
\begin{eqnarray}
a_{k+2}-2a_{k+1}+a_k
=
\frac{1}{2}g(   {\Delta t}       )^2
\left[(k+2)^2-2(k+1)^2+k^2\right]
=
g(     {\Delta t}         )^2.
\end{eqnarray}

Hence
\begin{eqnarray}{\Delta^2}z_k
=-g(     {\Delta t}     )^2,
\end{eqnarray}
which is independent of the value of $k$ {and is the discrete counterpart of the differential  relation $d(dz)=-g(dt)^2$}.

Recalling the physical relation between weight $P$, mass $m$, and gravitational acceleration $g$, namely
$
{P}/{m}=g,
$ {the discrete relation becomes }
\begin{eqnarray}
{\Delta^2}z=-\frac{P}{m}( {\Delta t})^2\Longrightarrow 
{m  {\Delta^2}z       =-P( {\Delta t} )^2},
\end{eqnarray}
{whose differential counterpart is $md(dz)=-P(dt)^2$, which} 
in modern mechanics, is expressed  as
\begin{eqnarray}
m\frac{d^2z}{dt^2}=-P,
\end{eqnarray}
what is now simply obtained by taking the second derivative of the  free fall trajectory
$z(t)=z(0)-\frac{1}{2}gt^2$.

The discrete relation
$
m  {\Delta^2}z  =-P(    {\Delta t}      )^2
$
makes explicit, however, the geometrical structure underlying Euler's formulation: namely, the constancy of the second spatial difference for equal time intervals. What is precisely the structure revealed by Galileo's law of odd numbers.

This provides a natural conceptual bridge between Galileo's synthetic geometry of accelerated motion and Euler's analytical formulation of mechanics. In Euler's notation, the corresponding second differential $d(dz)$ appears directly in his fundamental mechanical relation.
The present construction does not claim to reproduce Euler's original derivation \cite{Euler}. Rather, it offers a possible geometrical reconstruction of the structure underlying his introduction of $d(dz)$, by showing how it can naturally arise from Galileo's law of odd numbers and the constancy of the second spatial difference for equal time intervals. Whatever precise route Euler himself followed, the geometrical content of this relation would  surely have been  familiar within the Galilean tradition of reasoning about accelerated motion.

\subsection{Applications}
The conceptual power of the Equivalence Principle---using only pre-relativistic concepts and the constancy of the velocity of light---lies in its ability to transfer physical conclusions from an accelerated reference system to a static gravitational field.
Einstein's 1911 analysis provides a particularly striking illustration of this idea. Phenomena involving light can first be analyzed in an accelerated, gravity-free system, and then transferred, through the Equivalence Principle, to a gravitational field. 

For example, imagine a rocket in empty space. If a beam of light enters the rocket perpendicularly to its direction of motion, an observer standing on the rocket's floor will see the beam to follow a curved trajectory. Therefore, by applying the Equivalence Principle, one can infer that a uniform gravitational field--and, by extension, any gravitational field---must deflect light.

Another remarkable consequence follows from the same argument. Suppose that two light pulses are emitted from the ceiling towards the rocket's floor, separated by a time interval of one second, according to a clock carried by the observer in the rocket. Since the rocket is accelerating upward, the floor acquires a greater upward velocity during the time taken by the light to reach it. Thus, when the second pulse is emitted, the floor is already moving upward faster than it was before, when the first pulse was emitted. Since both pulses propagate at the same constant speed of light, the second pulse has less time to reach the floor than the first one. Consequently, the interval between the two arrivals is shorter than the one-second interval between their emissions.

By the Equivalence Principle, the same situation must be described by an observer at rest in a gravitational field. Two pulses emitted at a fixed height above the ground, one second apart according to a clock at that height, will therefore be received at the ground with a shorter time separation. The natural interpretation is that clocks at different gravitational potentials do not run at the same rate: the clock at lower potential records a shorter elapsed time between the two arrivals than the clock at higher potential, between the two emissions.

Thus, the Equivalence Principle leads to gravitational time dilation directly from the kinematics of light propagation, well before introducing any geometrical description of spacetime.

This illustrates the strong power of the Equivalence Principle: it is not merely an equivalence between two situations, but a rule that allows physical phenomena observed in one situation to be inferred in the other one.

Note that, in a purely Newtonian framework where light is assumed to consist of corpuscles, light would naturally be deflected like ordinary matter. This calculation was first performed by Johann Georg von Soldner \cite{Soldner} in 1801 ---although his work had little impact at that time, because the wave interpretation of light was then predominant. Soldner obtained the same result that Einstein derived  eleven decades later, in 1911 \cite{Einstein1911} using the Equivalence Principle (interpreting it as an indication that the speed of light varies under the influence of a gravitational field, combined with Huygens' principle). As was proven later, using the full theory of General Relativity, this result is exactly half of the correct physical deflection.

\begin{remark}
\label{rem:Soldner1911}
Retrospectively, the agreement between Einstein's 1911 prediction for the deflection of light and Soldner's earlier corpuscular calculation can be understood directly from the weak-field structure of Einstein's 1911 invariant. The two derivations, however, are physically and conceptually different: Soldner treats light as a massive corpuscle subject to Newtonian gravity, whereas Einstein derives the deflection from the gravitational dependence of the coordinate velocity of light and Huygens' principle. The following calculation shows why the two descriptions nevertheless lead to the same orbital equation.

Consider first a corpuscular particle of mass $m$ moving in the gravitational field of a spherically symmetric body of mass $M$. In polar coordinates, its Newtonian Lagrangian is
\begin{eqnarray}
    L_{\rm N}
    =\frac{1}{2}m\left(\dot r^2+r^2\dot\phi^2\right)
    +\frac{mMG}{r},
\end{eqnarray}
and the conserved angular momentum
\begin{eqnarray}
    J\equiv m r^2\dot\phi.
\end{eqnarray}
Choosing units such that the asymptotic velocity is $c=1$, the conserved energy for a particle arriving from infinity is
\begin{eqnarray}
    \frac{1}{2}m
    =
    \frac{1}{2}m\left(\dot r^2+\frac{J^2}{m^2r^2}\right)
    -\frac{mMG}{r}.
\end{eqnarray}
After dividing by $m$ and defining the specific angular momentum
\begin{eqnarray}
    \jmath\equiv\frac{J}{m},
\end{eqnarray}
this becomes
\begin{eqnarray}
    \frac{1}{2}
    =
    \frac{1}{2}\left(\dot r^2+\frac{\jmath^2}{r^2}\right)
    -\frac{MG}{r}.
\end{eqnarray}
Introducing $u=1/r$ and using
\begin{eqnarray}
    \dot r=-\jmath u',
\end{eqnarray}
where the prime denotes differentiation with respect to $\phi$, one obtains
\begin{eqnarray}
    (u')^2+u^2
    =
    \frac{1}{\jmath^2}
    +\frac{2MG}{\jmath^2}u,
\end{eqnarray}
and hence
\begin{eqnarray}
\label{Newton}
    u''+u=\frac{MG}{\jmath^2}.
\end{eqnarray}

Now consider Einstein's 1911 weak-field invariant {(the square of the infinitesimal proper time)}, in which the spatial sector remains Euclidean:
\begin{eqnarray}
    ds^2
    =
    \left(1-\frac{2MG}{r}\right)dt^2
    -\left(dr^2+r^2d\phi^2\right).
\end{eqnarray}
For a null trajectory, $ds^2=0$. Introducing an affine parameter $\lambda$, we may use the quadratic geodesic Lagrangian
\begin{eqnarray}
    L_{1911}
    =
    \frac12\left[
    \left(1-\frac{2MG}{r}\right)\dot t^2
    -\dot r^2-r^2\dot\phi^2
    \right],
\end{eqnarray}
where derivatives are with respect to $\lambda$.

Since the metric is independent of $t$ and $\phi$, the corresponding conserved quantities via the Euler-Lagrange equations are
\begin{eqnarray}
    \bar E
    =
    \left(1-\frac{2MG}{r}\right)\dot t,
    \qquad
    \bar J=r^2\dot\phi.
\end{eqnarray}
The null constraint gives, therefore,
\begin{eqnarray}
    \left(1-\frac{2MG}{r}\right)^{-1}\bar E^2
    -\dot r^2
    -\frac{\bar J^2}{r^2}
    =0.
\end{eqnarray}
To first order in the weak gravitational field,
\begin{eqnarray}
    \left(1-\frac{2MG}{r}\right)^{-1}
    \simeq
    1+\frac{2MG}{r},
\end{eqnarray}
so that
\begin{eqnarray}
    \dot r^2+\frac{\bar J^2}{r^2}
    =
    \bar E^2
    \left(1+\frac{2MG}{r}\right).
\end{eqnarray}
Using again $u=1/r$ and
\begin{eqnarray}
    \dot r=-\bar J u',
\end{eqnarray}
one obtains
\begin{eqnarray}
    (u')^2+u^2
    =
    \frac{\bar E^2}{\bar J^2}
    +\frac{2MG\bar E^2}{\bar J^2}u.
\end{eqnarray}
Thus, the orbital equation has precisely the same form as (\ref{Newton}), provided that the constants are normalized according to
\begin{eqnarray}
    \frac{\bar E}{\bar J}=\frac{1}{\jmath}.
\end{eqnarray}
Consequently,
\begin{eqnarray}
    u''+u=\frac{MG}{\jmath^2}.
\end{eqnarray}

The coincidence is, therefore, not a consequence of identical physical assumptions or identical derivations. Rather, it retrospectively follows from the fact that, to first order, Einstein's 1911 weak-field invariant modifies only the temporal sector,
\begin{eqnarray}
    g_{00}=1-\frac{2MG}{r},
    \qquad
    g_{ij}=-\delta_{ij},
\end{eqnarray}
while leaving the spatial sector Euclidean. The resulting null geodesic problem can consequently be reduced to an orbital equation of the same form as the Newtonian corpuscular problem, after a suitable normalization of the conserved quantities.

{
As we have noted above, Soldner's and Einstein's 1911 result provides only half of the full relativistic deflection. The missing contribution arises from the spatial sector, which in the present reconstruction is not introduced as an independent geometrical ansatz, but is determined by the requirement that the operational law of free fall be universal, in particular independent of the velocity of the test particle, as shown in \autoref{sectionV} and explicitly in remark \autoref{RemarkV.1}.

}

This observation should not be interpreted as a reconstruction of Einstein's 1911 derivation. Einstein did not obtain the deflection by treating light as a Newtonian corpuscle or by using geodesics. His original argument proceeds from the gravitational dependence of the coordinate velocity of light and Huygens' principle. The equivalence exhibited above is, therefore, a retrospective mathematical correspondence between the two descriptions, rather than evidence of a historical dependence of Einstein's calculation on Soldner's work.

Finally, the observational situation should be distinguished from the theoretical calculations discussed here. The 1914 German expedition to Crimea, organized to observe a solar eclipse and test the predicted deflection, was unable to obtain the intended measurements because of the outbreak of the First World War. The subsequent observations associated with the 1919 eclipse \cite{Dyson} were widely interpreted as supporting Einstein's general-relativistic prediction, which, unlike the 1911 result, includes the contribution of the spatial sector and gives approximately twice the deflection. And this matched perfectly with the observations, what {contributed to Einstein's public recognition.}
    \end{remark}

\section{Gravitational redshift}
In this section, we present a thought experiment designed to clarify the physical origin of the gravitational redshift. This approach is more robust than the standard, simplified version involving a single photon climbing a gravitational field, which often relies on the conceptually problematic assumption that a propagating quantum of light possesses an effective gravitational mass ($m = h\nu$) acted upon by a mechanical force. 
Instead, following the thermodynamic tradition of cyclic processes---famously adopted by Feynman \cite{Feynman1964}---we construct a global thought experiment where the gravitational redshift emerges as a rigorous and necessary consequence of strict energy conservation and the universal equivalence between mass and internal energy.

Consider two identical bodies, each with rest mass $m$\footnote{The bodies are assumed to be ideal test particles. Its own gravitational field is neglected, so that they probe, but do not modify, the prescribed external gravitational field.}. One is located on the Earth's surface, while the other is held at a height $z$ in a homogeneous gravitational field of strength $g$.

Initially, both bodies are in the same internal state. An external agent supplies an amount of energy $E_0$ to the lower body, increasing only its internal energy. According to Einstein's mass-energy relation, the mass of the lower body becomes
\begin{equation}
\bar{m}=m+{E_0}.
\end{equation}

The total energy of the system is, therefore,
\begin{equation}
E_{\rm tot}=2m+E_0+mgz,
\end{equation}
where the gravitational potential energy is measured relative to the lower body.

The lower body is then allowed to return to its original internal state by emitting all its excess internal energy exclusively as electromagnetic radiation. We consider an idealized situation in which the emitted radiation propagates without losses and is completely absorbed by the upper body.

Let $E_1$ denote the energy carried by the radiation when it reaches the upper body. After absorption, the upper body acquires the internal energy $E_1$, and its mass becomes
\begin{equation}
\tilde{m}=m+{E}_1.
\end{equation}
Since the additional internal energy contributes to the gravitational mass, the total energy of the final configuration is
\begin{equation}
E_{\rm tot}=2m+E_1+\left(m+{E_1}\right)gz.
\end{equation}

Now, as the system is isolated, total energy must be conserved. Equating the initial and final energies yields
\begin{equation}
E_0=E_1\left(1+{gz}\right),
\end{equation}
which coincides with the result obtained by Einstein in 1911 using the equivalence between an accelerated system and one at rest under the action of a uniform gravitational field (the Equivalence Principle) \cite{Einstein1911}.

Finally, adopting Einstein's 1905 heuristic hypothesis that electromagnetic radiation consists 
of energy quanta\footnote{Although proposed by Einstein in 1905, the light-quantum hypothesis remained controversial for nearly two decades. Only after the experimental confirmation of the Compton effect did it become widely accepted.} \cite{Einstein1905} satisfying
\begin{equation}
E_1=h\nu_1,\qquad E_0=h\nu_0,
\end{equation}
one immediately finds
\begin{equation}
\nu_0 = \nu_1 \left(1+{gz}\right),
\end{equation}
or, for weak fields,
\begin{equation}
\nu_1=\nu_0\left(1-{gz}\right).
\end{equation}

Therefore, this derivation demonstrates that the gravitational redshift follows directly from energy conservation together with the equivalence between mass and energy, without strictly requiring the invocation of accelerated frames or the Equivalence Principle.

Most importantly, this result shows that gravity affects the rate at which proper time flows. Since the frequency of any physical clock is shifted by the gravitational field, clocks located at different gravitational potentials cannot accumulate proper time at the same rate. Consequently, the proper time elapsed between two events depends on the gravitational potential experienced by the clock.

This phenomenon is not a mere theoretical deduction, but indeed one of the most rigorously tested empirical facts in modern physics. The shift in frequency due to a gravitational field was first confirmed by the Pound-Rebka experiment \cite{Pound1960}, which measured the gravitational redshift of gamma rays in a terrestrial laboratory. Subsequently, the macroscopic divergence in the accumulation of proper time was explicitly demonstrated by the Hafele-Keating experiment \cite{Hafele1972}, which compared macroscopic atomic clocks transported in commercial airliners against ground-based standards. More recently, advances in optical atomic clocks have achieved such extreme precision that this differential rate of proper time can be measured for altitude differences as small as a fraction of a meter \cite{Chou2010, Bothwell2022}. Such experimental milestones demonstrate that the gravitational modulation of proper time is a direct physical reality, fully supporting an operational approach based on physical clocks. 

\subsection{Problems with the Equivalence Principle and Geometric Reification}
In our derivation of the gravitational redshift, we deliberately bypassed Einstein's 1911 approach based on the Equivalence Principle (EP) \cite{Einstein1911}.
The reason is that the EP is frequently misconstrued as a literal physical identity rather than a heuristic stepping stone. Historically, the principle was invoked to extrapolate physical results from a uniformly accelerated reference frame (in flat spacetime) to a frame at rest in a uniform gravitational field. However, this analogy has profound limitations. Today,
within the framework of modern differential geometry, we understand that a genuine gravitational field involves intrinsic spacetime curvature (a non-zero Riemann tensor), whereas a uniformly accelerated frame is merely a coordinate transformation within flat Minkowski spacetime. For this reason, many modern physicists echo J.L. Synge's famous remark \cite{Synge1960} that the Equivalence Principle is a historical relic that should now be "buried with honours."

The heuristic nature of the Equivalence Principle also helps explain why Einstein's route toward the final theory proceeded in several conceptually distinct stages. Initially, in 1912 \cite{Einstein1912}, the application of the EP (through the elevator thought experiment) led him to conclude that a gravitational field alters the rate at which clocks at rest at different heights run. Consequently, his first treatment described a static gravitational field solely through a modification of the temporal component $g_{00}$
of the invariant $ds^2$, since for bodies at rest with respect to the observer the proper-time element reduces to\begin{eqnarray}ds=\sqrt{g_{00}}dt.\end{eqnarray}
This approach soon revealed important difficulties. For example, it failed to provide consistent equations of motion, predicting only half of the observed deflection of light. Einstein therefore realized in 1912 that gravity must modify the entire structure of the invariant $ds$, not merely its temporal sector.

To justify why gravity should also modify the spatial sector of the invariant, Einstein turned to a second heuristic argument: the Ehrenfest paradox \cite{Ehrenfest1909}. Considering a rotating disc, he argued that inertial observers applying Lorentz contraction would conclude that the geometry of the disc is non-Euclidean. He then invoked the Equivalence Principle to argue that if an accelerated frame exhibits non-Euclidean spatial geometry, a gravitational field should do so as well. This two-step reasoning---using the elevator to motivate the temporal sector and the rotating disc to motivate the spatial sector---eventually led Einstein toward a fully geometric formulation of gravitation.

Why did Einstein feel compelled to proceed in this convoluted manner? The root of the problem lies in how he conceptualized the invariant $ds$ between 1907 and 1916. Throughout this crucial period,
unlike Minkowski, Einstein did not make the operational interpretation of $ds$ as the differential of proper time the central organizing principle of his construction.
 Instead, heavily influenced by Marcel Grossmann and the legacy of 19th-century mathematics, he viewed $ds$ through the lens of Gaussian surface geometry. In his 1913 \textit{Entwurf} paper, he called $ds$ the "natural distance," and in his definitive 1916 paper, he referred to it, borrowing the name from  Gauss theory of surfaces,  as the "linear element" (line element).
This historical development naturally emphasized the geometric interpretation of $ds$, while the operational interpretation in terms of proper time remained in the background. From the viewpoint adopted in this work, this shift led to what may be regarded as a reification of the mathematical formalism.

Things become remarkably simpler, and 
conceptually more transparent, when one abandons this geometric reification and accepts the strictly operational view. As Hermann Minkowski established in 1908, the invariant $ds_{\rm Min} = \sqrt{dt^2 - dx^2 - dy^2 - dz^2}$ (in the coordinates of an idealized inertial observer without gravity) is nothing more than the infinitesimal proper time. The integral $s = \int_{\gamma} ds_{\rm Min}$ is quite simply the time elapsed on a physical clock moving along the worldline $\gamma$.

From this operational viewpoint, gravity is naturally understood as modifying the coefficients appearing in the invariant proper-time element $ds^2$, instead of leaving them equal to those appearing in the Minkowskian expression of the invariant proper-time element.
 This viewpoint eliminates the need to motivate separately the temporal and spatial sectors of the invariant through different heuristic arguments, namely, elevators and spinning discs.
   Once we recognize $ds$ operationally as the infinitesimal proper time, the modification of the proper-time structure by gravity
    is understood directly as a unified modulation of temporal rhythms, thereby avoiding the conceptual route that historically led Einstein to rely on separate heuristic constructions.

\subsubsection{The operational interpretation of the rotating-disk experiment
}
Consider two observers on a disk initially at rest, equipped with two bags containing identical infinitesimal rods of proper length $L_0$, labeled $A$ and $B$. Using $N_{A, \text{cir}}$ rods from bag $A$,
they construct on the disk a circular concentric ring, with  perimeter
 $N_{A, \text{cir}} L_0$, and they weld the ends of the rods together, so that the circumference forms a  welded ring.

They also use $N_{A, \text{dia}}$ rods to measure the diameter of this circle, obtaining $N_{A, \text{dia}} L_0$. Consequently, the ratio between the perimeter and the diameter, in rest conditions, is
\begin{equation}
\frac{N_{A, \text{cir}}}{N_{A, \text{dia}}} = \pi.
\end{equation}
Next, one of the observers steps off the disk and activates the motor that sets the disk in rotation. The angular velocity is kept sufficiently small so that the mechanical stress on the welded rods does not exceed their breaking threshold. An observer outside the disk continues to 
see an intact, continuous circumference.

Then, the observer remaining on the rotating disk takes the unconstrained rods from bag $B$, and lays them along the circumference. 
The crucial question is: how many rods from bag B will the rotating observer need, in order to measure the complete circumference?

{In the external inertial description, the free rods from bag \(B\) are Lorentz-contracted along their tangential direction. If the circumference is reconstructed by assigning this contracted length to each infinitesimal measuring rod, more rods are required to cover the circumference, so that}
$N_{B, \text{cir}} > N_{A, \text{cir}}$. 
Crucially, as Einstein established in 1905, this contraction is purely operational: it strictly arises from the spatio-temporal measurement protocol employed by an observer in relative motion. It is not the result of a physical compression, where molecules are squeezed together by internal forces, as hypothesized by FitzGerald or Lorentz. In the proper frame of each free rod, its length remains $L_0$; 
the contracted value is a frame-dependent result of the spatio-temporal measurement protocol used by the moving observer.

Furthermore, the external observer observes that the rods placed along the diameter do not undergo contraction, as their velocity is strictly orthogonal to their orientation. Thus, $N_{B, \text{dia}} = N_{A, \text{dia}}$. Consequently, within the external observer's map, the ratio of the perimeter to the diameter is given by:
\begin{equation}
\frac{N_{B,\text{cir}}}{N_{B,\text{dia}}} = \frac{N_{B,\text{cir}}}{N_{A,\text{dia}}} > \frac{N_{A,\text{cir}}}{N_{A,\text{dia}}} = \pi.
\end{equation}
From this operational ratio, one might conclude that, within the external observer's map, the geometry on the disk is non-Euclidean.

{
The present thought experiment makes explicit a distinction that is often left implicit in operational discussions of the rotating disk: the distinction between the material circumference (bag $A$) on which the measurement is performed and the freely movable rods (bag $B$) used as operational measuring standards.}

At this point, however, a distinct conceptual issue emerges. The operational argument determines how the free rods from bag $B$ are represented in the external observer's map, ultimately yielding a non-Euclidean ratio
{within the external observer's reconstruction.} It does not, however, determine the physical state of the welded rods forming the material circumference. 

Importantly, we do not assume any particular relativistic or elastic deformation of the welded rods. Their physical state after the onset of rotation is not determined by the kinematical Lorentz transformation alone. The welded rods constitute a mechanically constrained system, and their actual deformation, if any, would require a dynamical treatment of the material, including its elastic properties and the stresses generated by the rotational constraint.

It would therefore be conceptually inappropriate to combine a Lorentz-contracted length assigned by an external inertial observer with an elastic deformation of the welded ring and treat the two as independent corrections to the same geometrical quantity. The former belongs to the kinematical description of measurement, whereas the latter belongs to the physical dynamics of the material system.

For this reason, the present construction deliberately leaves the physical state of the welded rods unspecified. This is not a deficiency of the argument, but part of its purpose: the operational result obtained from the free rods in bag $B$ can be discussed independently of any assumption about the mechanical response of the circumference of the material formed by the rods in bag $A$.

Since Lorentz contraction remains far below experimental resolution for angular velocities compatible with the structural integrity of the welded ring, this thought experiment is strictly conceptual. Its purpose is not to invalidate Einstein's kinematic calculus but to untangle two logically distinct questions: (i) how an observer operationally measures lengths using free rods, and (ii) what physical state is attributed to the material object on which those measurements are performed. 

Conflating these two levels risks reifying an operationally defined geometrical relation into an ontological statement about space itself.
The experiment {shows} that the non-Euclidean ratio obtained from the operational procedure does not, by itself, determine the ontological status of geometry. Whether it is interpreted as a property of the observer's operational map or as a property of the physical geometry itself, belongs to the level of interpretation rather than to the operational content of the experiment.

In fact, the rotating-disk experiment illustrates that operational geometry and material geometry need not be identified. The operational protocol determines how lengths are assigned within an observer's map, whereas the physical state of the material object belongs to a different level of description.

A closely related operational point of view can be found in Tartaglia [28], who emphasizes the {importance of specifying the measurement procedure when discussing lengths on rotating platforms. In particular, for an observer co-rotating with the disk, the circumference is measured by counting rods that are at rest with respect to the observer; these rods therefore retain their proper length and do not undergo Lorentz contraction in that observer's frame. Tartaglia argues that, under this operational procedure, the circumference-to-diameter ratio remains \(2\pi\). The result illustrates that the geometrical interpretation assigned to a rotating system depends essentially on how lengths are operationally defined and measured.

}

Unlike Einstein's operational construction in Special Relativity, which relies on a single inertial observer equipped with synchronized clocks, the rotating-disk argument {considered here distinguishes between the global reconstruction 
adopted in the external inertial description
and the direct measurement protocol available to the observer co-rotating with the disk. In the former description, the circumference is reconstructed by applying the local Lorentz contraction to infinitesimal measuring rods distributed along the non-inertial trajectory. The resulting circumference is therefore a theoretical reconstruction rather than the outcome of a single direct measurement.} In contrast, Tartaglia explicitly defines the measurement protocol adopted by {the rotating observer: the observer places rods that remain at rest in his reference frame directly along the circumference, thereby recovering the Euclidean circumference-to-diameter ratio within his operational procedure.

}

The operational calculation is therefore well defined. What remains open is not the calculation itself, but its ontological interpretation. The rotating-disk experiment establishes an operational relation between measured lengths; it does not, by itself, determine whether the resulting non-Euclidean relation should be regarded as a property of physical space, as a property of the observer's operational representation, or as an effective description arising from the relativistic kinematics of the rotating system. The distinction between these possibilities belongs to the level of interpretation rather than to the operational content of the measurement.

\subsubsection{On the practical feasibility of the rotating-disk experiment}
It is instructive to estimate the practical feasibility of the rotating-disk thought experiment. Consider a circular ring of radius $R$ approximated by a regular polygon consisting of $N$ identical rods. The proper length of each rod is
\begin{equation}
L_0
=
2R\sin\left(\frac{\pi}{N}\right)
\simeq
\frac{2\pi R}{N},
\end{equation}
that is, the circumference divided into $N$ equal segments.

As a concrete example, let
$
R=3~\mathrm{km},
$
and suppose that the circumference is constructed from
$
N=2\times10^6
$
identical rods. Then, 
each rod has proper length
\begin{equation}
L_0
\simeq
\frac{2\pi(3\times10^3)}{2\times10^6}\,\mathrm{m}
\simeq
9.4\times10^{-3}\,\mathrm{m}
\approx
1~\mathrm{cm}.
\end{equation}

Assume now that the disk rotates at one revolution per second. The tangential velocity at the rim is
\begin{equation}
v
=
2\pi R
\simeq
1.88\times10^4\,\mathrm{m/s}
\approx
19~\mathrm{km/s}.
\end{equation}
Although this angular velocity is already extremely large, from a mechanical point of view, it still satisfies
$
\frac{v}{c}\ll1,
$, where here we take explicitly $c\cong 3\times 10^5$ km/s, 
so that
\begin{equation}
\sqrt{1-\frac{v^2}{c^2}}
\simeq
1-\frac{v^2}{2c^2},
\end{equation}
with
$
\frac{v^2}{2c^2}
\simeq
2\times10^{-9}.
$

The total Lorentz contraction of the entire circumference,
$
L=2\pi R,$
is therefore
\begin{equation}
\Delta L
=
L\left(1-\sqrt{1-\frac{v^2}{c^2}}\right)
\simeq
L\frac{v^2}{2c^2}
\simeq
8\times10^{-5}\,\mathrm{m},
\end{equation}
that is,
\begin{equation}
\Delta L
\simeq
0.08~\mathrm{mm}.
\end{equation}

This contraction is less than one hundredth of the length of a single rod, whose proper length is approximately $1\,\mathrm{cm}$. Consequently, both bags $A$ and $B$ would require essentially the same number of rods to cover the circumference.

If the angular velocity is reduced, in order to ensure that the welded ring remains well within its elastic regime, the Lorentz contraction decreases quadratically with $v$, becoming even smaller. Therefore, although the rotating-disk argument is conceptually illuminating, its direct experimental realization is practically unattainable. The operational difference between the two sets of rods remains far beyond any realistic experimental resolution in the regime where the material ring preserves its structural integrity.

\section{Conatus and Potentia}
The present section adopts the methodology of classical natural philosophy: before constructing the mathematical formalism, one first asks what the natural law governing the motion of a  {material} free body is.

To answer this question, we recover the old notion of \textit{conatus}, a concept with a long and distinguished history in natural philosophy \cite{Jammer1957, Wolfson1934}.
Rooted in the Stoic notion of \textit{horme}—the intrinsic striving of living beings to preserve themselves—and translated into Latin as \textit{conatus} by Cicero, the concept originally referred to a metaphysical tendency toward persistence in existence. However, during the seventeenth century the term acquired a different meaning within natural philosophy. Hobbes interpreted it as an infinitesimal beginning of motion
\cite{Hobbes1655}, Descartes associated it with the tendency of bodies to preserve their state of motion or rest
\cite{Descartes}, and Newton ultimately reformulated it as the \textit{vis insita} or \textit{vis inertiae} \cite{Newton1687}, 
the innate property of matter by which every body resists any change in its state of rest or uniform motion.

Newton's formulation, unlike Descartes', 
has a remarkable internal logical coherence because it is built upon the notions of "absolute space" and "absolute time". Motion is not defined merely as the changing relation between bodies but with respect to a fixed and immutable background. A body tends to remain at rest or in uniform rectilinear motion relative to this absolute space, and inertial observers are precisely those whose motion is uniform with respect to it. Newton's first law, therefore, does not merely describe how bodies move; it also identifies the privileged reference systems from which the laws of mechanics are to be formulated.

Descartes adopted a fundamentally different view. For him, space was inseparable from matter itself: extension constituted the very essence of bodies, and empty space had no physical meaning. Motion was therefore purely relational, describing only the changing configuration of material bodies with respect to one another. This raises a conceptual difficulty. His idea of \textit{conatus} indicates that every body tends to preserve its state of motion or rest, but if all motion is purely relative, the natural question immediately arises: uniform with respect to what? Newton's introduction of absolute space provides a precise answer to this question and removes the ambiguity that remains in the Cartesian formulation.

Within Newtonian mechanics, these "inertial" observers occupy a privileged position. They constitute the natural reference systems from which the laws of motion are formulated, while all other observers are obtained by transformation from them.

This picture changes profoundly, once gravitation is introduced. Because gravitation acts universally and cannot be screened, no observer can ever be perfectly isolated from its influence. Consequently, the privileged status of inertial observers loses its ontological meaning. Absolute space remains part of the Newtonian construction, but  privileged observers become idealizations that can only be approximated operationally.

The difficulty is therefore not mathematical but conceptual. Newtonian mechanics requires the laws to be formulated with respect to observers that no physical laboratory can ever exactly realize.

In contrast, D'Alembert's principle removes this asymmetry. Rather than assigning a privileged status to inertial observers, it formulates mechanics as a balance law that can be written in any reference frame, provided the appropriate inertial contribution is included. What remains invariant are not the involved forces, but the balance itself.

Baruch Spinoza subsequently recaptured the broader essence of the old tradition \cite{Spinoza}. While his actual physics remained largely Cartesian, in his metaphysical magnum opus, the \textit{Ethics}, he transcended purely mechanical restrictions to elevate \textit{conatus} to a universal ontological principle: the immanent, driving endeavour of every individual thing to persevere in its own being (\textit{in suo esse perseverare conatur}).

We here adopt this broader Spinozian conceptual framework as an interpretive lens, rather than a strictly mechanical Cartesian-Newtonian one. Although originally conceived as a metaphysical doctrine, Spinoza's ontological \textit{conatus} offers a striking conceptual parallel when projected onto modern relativistic physics. If the physical "existence" of a fundamental {massive} particle is fundamentally quantified by the ticking of its own clock, then its innate tendency to persevere in existence  seamlessly translates into an endeavour to maximize, {at least locally}, its own proper time---{an} invariant quantity intrinsically associated with its history. A free-falling body follows a worldline for which the elapsed proper time between two neighbouring events is extremal (specifically, a maximum for timelike trajectories). In this sense, proper time is not merely a geometrical invariant but the operational manifestation of the particle's \textit{conatus}: left undisturbed by external non-gravitational forces, 
every body spontaneously evolves along {a locally maximizing worldline of its own internal time.}

The second crucial Spinozian notion is \textit{potentia}.\footnote{
In his \textit{Ethics} (III, Prop.~7 and IV, Prop.~4), Spinoza identifies the \textit{conatus} of a singular thing as a localized and finite expression of the infinite \textit{potentia} of Nature. In our physical map, this relation is beautifully preserved: the \textit{conatus} of the test particle---its intrinsic tendency to {locally} maximize its own proper time---navigates and responds to the \textit{potentia} of the gravitational field, which is the global distribution of matter-energy that rewrites local temporal rhythms, namely the capacity of a substance to act upon and influence another.} In the present interpretation, gravitation plays precisely this role. In the present map, matter does not exert its influence by generating an external force acting at a distance in the traditional Newtonian sense; rather, it acts by dynamically modulating the rate of proper time across spacetime. As demonstrated by the thought experiment in the previous section, the presence of matter alters the physical ticking rate of local clocks situated at different positions in the field. Consequently, the free motion of test particles is modified not because they are mechanically pushed or pulled, but because the non-uniform temporal landscape governing their proper time has been reshaped by the surrounding mass distribution. From this perspective, gravitation is reinterpreted as the physical realization of \textit{potentia}: matter acts upon matter by shaping the local operational rhythm of time. Free fall then emerges as the natural and inevitable consequence of the interplay between \textit{conatus}---the intrinsic tendency of {material objects} to {locally maximize the number of ticks of their internal clock}---and \textit{potentia}---the structural influence of matter on 
the proper-time landscape experienced by free {massive} particles. 

{
The \textit{conatus} principle is therefore restricted to massive material particles. For such particles, free motion is characterized by the stationarity of their experienced proper time, which, under the appropriate local conditions, corresponds to a local maximum. Light requires a different treatment. In Special Relativity, the constancy of the speed of light, 
using the notation ${\bf r}=(x,y,z)$ is
\(d{\bf r}/dt=1\) in units \(c=1\),  and implies
$$
ds^2=dt^2-d{\bf r}^2=0,
$$
along every light trajectory. Hence, the proper time experienced along the propagation of light vanishes identically, and the \textit{conatus} principle cannot be used to determine its trajectory. In the presence of gravitation, we retain the same operational requirement: light propagation is characterized by vanishing proper time, \(ds=0\). {The condition $ds^2=0$ determines the null structure, but the actual propagation of freely moving light rays is determined by the corresponding null geodesics.}
Thus, although the \textit{conatus} principle applies specifically to massive material particles, the operational structure of time that it helps reconstruct also constrains the propagation of light through the condition of vanishing proper time.

}

From this operational perspective, the physical significance of the Riemann tensor is no longer understood primarily as the intrinsic curvature of a substantival spacetime fabric, but rather as the mathematical manifestation of the non-uniformity within the proper-time landscape. Just as Riemann originally introduced curvature to quantify the local departure of a metric from Euclidean geometry \cite{Riemann1867}, the Riemann tensor is interpreted here as characterizing the departure of the local proper-time structure from the uniform temporal baseline of Minkowski spacetime.

It is precisely the presence of  second-order gradients(tildal forces)—the fact that the \textit{potentia} of surrounding matter acts non-uniformly across space—that forces neighbor free-falling particles to deviate from one another. In this relational framework, the Riemann tensor does not describe the properties of an independent geometric arena. Instead, it serves as the mathematical synthesis of the relative deviation of neighbor worldlines, a deviation arising directly from the spatial variations that matter imprints upon the flow of local proper time.

Ultimately, a fundamental conceptual distinction emerges. Riemann's original formulation was naturally expressed as a geometry of infinitesimal spatial distances, where the metric defines the elementary notion of length. General Relativity, viewed operationally, begins instead with clocks rather than rods: its primary observable is the infinitesimal proper time experienced by physical systems. Once the invariance of the speed of light is postulated, spatial distances cease to be primitive entities and are reconstructed operationally from light signals and clock comparisons. 
The mathematical object is essentially the same; what changes is its operational interpretation: rather than being introduced as an abstract measure of distance, it encodes the local flow of proper time.

Yet, this operational reconstruction remains remarkably faithful to Riemann's methodological vision. In his \textit{Habilitationsvortrag} \cite{Riemann1867}, Riemann argued that the metric relations of a continuous manifold could not be derived from geometry alone but must ultimately originate from physics itself. 
Therefore, he did not regard the metric as a self-subsisting geometric object but as the mathematical expression of deeper physical relations. In this respect, the present work may be viewed as realizing, within the relativistic framework, precisely the methodological programm that Riemann left open: geometry does not explain physics; physics determines the metric structure that geometry subsequently formalizes.

Finally, it is crucial to emphasize that the concepts of \textit{conatus} and \textit{potentia} are not introduced here as mere historical metaphors, but rather as the two fundamental dynamical postulates from which the present formulation of gravitation is developed. Importantly, however, the framework proposed here is epistemological rather than ontological. It does not claim that nature is literally governed by \textit{conatus} and \textit{potentia}; but these concepts provide the conceptual map through which the dynamics of gravitation can be consistently described. 

In this sense, it shares the same epistemological status as Newton's framework of gravitational forces: it is a highly effective representation of interactions, without thereby committing us to an account of their ultimate physical nature. Echoing Newton's famous dictum \textit{hypotheses non fingo}, we do not attempt to formulate hypotheses concerning the ultimate physical essence or intrinsic cause of this \textit{potentia}. We simply acknowledge its existence as a relational principle and develop the mathematical formulation that describes how this gravitational field—our specific representation of mass interactions—modifies the proper time of test particles. By shifting from the language of external forces to the language of modified proper time, we are not changing the territory of reality, but just offering a different, dynamically consistent chart with which to navigate it. The territory remains the same; only the map is changed.

\subsection{The Variational Origin of the Equality of Inertial and Gravitational Mass}

It is essential to clarify the role of mass ($m$) within this variational architecture. We do not treat mass as an artificial parameter or a fiction; it remains a fundamental physical property of matter, identified by Special Relativity as the total rest-energy of the system ($E = m$). However, because our fundamental dynamical principle {requires the elapsed proper time along a free worldline to be stationary} ($\delta \int ds = 0$), the resulting equation of motion---the geodesic equation---is  purely kinematical and strictly
independent of the particle's own mass.

The classical map of forces emerges only \textit{a posteriori} as a macroscopic translation designed to recover Newtonian mechanics. To express this temporal optimization in terms of a mechanical force, one must explicitly multiply both sides of the kinematical equation by the particle's rest mass, $m$. On the left-hand side, this mass scales the acceleration, operationally defining what classical physics termed \textit{inertial mass} ($m_{\rm i}$). On the right-hand side, the same mass scales the gradient of the gravitational potential, defining what was termed \textit{gravitational mass} ($m_{\rm g}$).

Consequently, within this framework, the identity $m_{\rm i} = m_{\rm g}$ ceases to be an unexplained empirical coincidence or an independent axiom of nature. Because both "forces" are generated by multiplying a single, underlying kinematical equation by the exact same parameter—the physical rest-energy of the body—the apparent duality of masses dissolves. The mystery is revealed to be an artifact of the Newtonian map: there are not two distinct properties that fortuitously share the same value, but rather a single rest-energy interacting with a modified landscape of physical time. Nothing physical has disappeared; only the conceptual duplication introduced by the Newtonian force language is eliminated.

Finally, this variational derivation bypasses the traditional need to invoke Einstein's Equivalence Principle as an independent postulate to justify why both masses share the same nature. Indeed, a critical review of the Equivalence Principle reveals that equality $m_{\rm i} = m_{\rm g}$ was already implicitly encoded in its formulation (where constant acceleration in a uniform gravitational field implies the numerical identity of both masses). By contrast, within the proper-time map developed here, founded on the principles of \textit{conatus} and \textit{potentia}, this identity is not assumed geometrically from the outset, but emerges as a {direct} consequence of expressing temporal optimization in the language of classical mechanics.

\section{Gravitation for weak fields}\label{sectionV}

Having established in the previous sections that gravity acts by modifying the flow of proper time, we are now in a position to formulate the central program of this essay. 
{Our goal} is to determine how the \textit{potentia} generated by material sources alters the proper-time structure experienced by a test particle, and how the particle's \textit{conatus} responds by {tending toward the worldline of {locally} maximal proper time, as expressed variationally by the stationarity condition $\delta\int ds=0$} \cite{Haro2026}. The construction proceeds in the following steps:

\begin{enumerate}
    \item \textbf{Static Matter:} 
    {We formulate the universal law of Galilean free fall in terms of the proper time carried by the freely falling particle,}
requiring it to be independent of the particle's velocity, to find how static masses modify proper time.
    \item \textbf{Moving Matter:} We apply Lorentz transformations to the linearized static invariant to deduce, within the weak-field limit, how moving masses alter the proper-time structure.
\end{enumerate}

{
Regarding the first point, we formulate the universality of Galilean free fall in terms of the proper time carried by the freely falling massive object. 
Specifically, we postulate that, in a uniform gravitational field, the particle's acceleration with respect to its proper time obeys
}

\begin{equation}\label{eq-freefall}
\frac{d^2z}{ds^2}=-g.
\end{equation}

{
This is not meant as a mere reparametrization of the Newtonian law $d^2z/dt^2=-g$.
Rather, it is an additional dynamical postulate of the present framework, motivated by the identification of the inertial temporal parameter with the proper time carried by the particle itself.

In fact, the main  motivation for this choice of temporal parameter comes from the physical meaning assigned to the inertial term in D’Alembert’s principle.

}

As discussed in \autoref{sec-PE}, the balance law describing a freely
falling body, as written by an observer at rest on the Earth's surface,
is
\begin{equation}
F_{\rm i}-mg=0,
\end{equation}
where
\begin{eqnarray}
F_{\rm i}\equiv -m\frac{d^2z}{ds^2}.
\end{eqnarray}

The inertial contribution depends on the observer: an observer at rest
on the Earth's surface attributes an inertial force $mg$ to the falling
body, whereas an observer freely falling with the body attributes no
such inertial contribution. What changes between these descriptions is
therefore the acceleration assigned to the body, not the physical
system itself.

This distinction suggests that the temporal parameter entering in the
inertial term should not be tied to any particular external observer.
The particle itself carries a clock along its worldline, and the proper
time measured by this clock is independent of which observer describes
the motion. It therefore provides a natural temporal parameter for an
inertial quantity that is to be associated with the particle itself.

In Newtonian mechanics this distinction is immaterial because all
observers share the same absolute time. In a relativistic theory,
however, no universal temporal parameter exists. Observers located at
different gravitational potentials may assign different elapsed
 times to the same physical process. For example, consider
an object released from rest at the top of a tower with observers at
the base and at the point of release. Their stationary clocks do not
accumulate time at the same rate because they are located at different
gravitational potentials. They therefore provide different temporal
descriptions of the same fall.

However, the falling object carries its own clock. 
{Along its worldline, the particle carries its own clock, which measures the proper time $s$,}
and the equation
\begin{eqnarray}
\frac{d^2z}{ds^2}=-g,
\end{eqnarray}
describes the evolution of its height with respect to this intrinsic
temporal parameter.
The spatial quantity $z$ denotes the height of the
particle above the Earth's surface and can therefore be used as a
common spatial variable in the descriptions given by {stationary} observers at
different locations. What is not common to these observers is the time used to describe the evolution of $z$.

Thus, the use of proper time is not an arbitrary choice of
parametrization. It expresses the requirement that the temporal
measure entering the inertial contribution be attached to the physical
system rather than to the particular observer describing it.

We now seek the invariant line element whose extremal proper-time principle reproduces exactly the free-fall law (\ref{eq-freefall}), assuming the initial condition
 $\frac{dx}{ds}
=\frac{dy}{ds}=0$.

As we will now show,
{under the assumption of spatial isotropy}, 
this  leads to the  invariant
\begin{equation}\label{eq-metric-static}
ds^2 = \left(1+{2gz}\right)dt^2 - \left(1+{2gz}\right)^{-1} d{\bf r}\cdot d{\bf r},
\end{equation}
where, {once again},  we have used units where the velocity of light is $1$ and the notation ${\bf r}=(x,y,z)$.

{With the spatially isotropic ansatz, the invariant takes the form:}
\begin{eqnarray}
    ds^2=A(z)dt^2-B(z)d{\bf r}\cdot d{\bf r},
\end{eqnarray}
and  calculate {the dynamics assuming that the proper time is stationary, i.e.,} $\delta \int ds=0$ to obtain, in terms of proper time,
\begin{eqnarray}
    \frac{d^2 z}{ds^2}=-\frac{A'+B'\left(\frac{dz}{dt}\right)^2}{2B\left(A-B\left(\frac{dz}{dt}\right)^2\right)}
    =-\frac{A'+(AB)'\left(\frac{dz}{ds}\right)^2}{2AB},
\end{eqnarray}
where the prime denotes the derivative with respect to the variable $z$.

The remarkable point is that, {within this ansatz,} demanding the equation of motion to be completely independent of the particle's speed uniquely imposes the condition
$B=a/A$, where $a$ is a constant. However, since at $z=0$ the gravitational potential vanishes, the invariant must be the Minkowski one what imposes $a=1$.

Then,
\begin{eqnarray}
     \frac{d^2 z}{ds^2}=-A'/2,
\end{eqnarray}
an to obtain the free-fall law (\ref{eq-freefall})
 we have to choose $A(z)=k+2gz$.
{Moreover}, since at \(z=0\) the potential vanishes, {we require the invariant to reduce to the Minkowski one}, which implies \(k=1\), that is,
\begin{eqnarray}
    A(z)=1+2gz.
\end{eqnarray}

It is worth
{emphasizing that no gravitational geometry has been postulated} at this stage. The invariant is obtained 
from the dynamical law of free fall, together with the variational principle of extremal proper time, {within the spatially isotropic ansatz introduced above. Geometry enters only afterwards,} as the mathematical interpretation of the resulting invariant.

\begin{remark}\label{RemarkV.1}
It is instructive to observe that the exterior Schwarzschild invariant can be reconstructed, {within a static spherically symmetric ansatz}, directly from the Newtonian law of free fall written in terms of proper time.

Consider a particle in radial free fall in a spherically symmetric gravitational potential $\Phi(r)$ and assume that its motion satisfies
\begin{equation}
    \frac{d^2r}{ds^2}=-\Phi'(r),
\end{equation}
where the prime denotes differentiation with respect to $r$.
{Within the static spherically symmetric ansatz, let the invariant proper-time element take the form
}
\begin{equation}
    ds^2=A(r)dt^2-B(r)dr^2-r^2d\Omega^2.
\end{equation}

Requiring the above equation of motion to hold {within this ansatz} determines the metric coefficients as
\begin{equation}
    A(r)=B^{-1}(r)=1+2\Phi(r).
\end{equation}

For the Newtonian potential
\begin{equation}
    \Phi(r)=-\frac{GM}{r},
\end{equation}
one immediately obtains
\begin{equation}
    A(r)=B^{-1}(r)=1-\frac{2GM}{r},
\end{equation}
{which reproduces the standard exterior Schwarzschild line element, within the static, spherically symmetric ansatz adopted above and with \(r\) interpreted as the areal radius.
}

\end{remark}

The second step, namely to extend this result to moving sources, requires careful epistemological clarification. To {determine} how moving masses affect proper time, we restrict {the construction} to the weak-field regime, retaining only terms linear in the gravitational potential.

The rationale for this restriction is physical. {In the present construction, Lorentz transformations are used as kinematical operations on the Minkowski background to describe the motion of localized material sources. At linear order, the gravitational field admits the superposition of independent contributions from such sources, so the static field generated by each mass element can be Lorentz-transformed and subsequently combined to describe a moving distribution.

Beyond the linear approximation, however, this procedure is no longer justified. Although a Lorentz transformation can in principle be applied to the complete line element, the nonlinear terms cannot be interpreted as a simple superposition of independently boosted elementary fields. Products of the gravitational perturbation encode the nonlinear self-interaction of the gravitational field itself. The field therefore contributes to its own effective source, and its dynamics can no longer be separated from the kinematics of the material sources by a simple Lorentz boost.}

For this reason, the Lorentz transformation is applied {here} exclusively to the linearized static invariant. {This isolates the contribution associated with the kinematic state of the material sources while consistently neglecting gravitational self-interaction, as required by the weak-field approximation. The nonlinear completion of the construction, in which the gravitational field itself contributes to the source of the gravitational field, is addressed subsequently.}

Accordingly, within this linear regime, we assume that the motion of the material sources is purely kinematic and unaffected by mutual gravitational interactions. 
This ensures that the {boost is used only to describe the kinematics of the sources and does not incorporate second-order gravitational self-interaction into the linearized construction.}

{The resulting linearized field equations coincide with those of General Relativity in harmonic gauge}
(details will appear in \cite{Haro2026a}), that is, writing 
\begin{equation}
ds^2=g_{\mu\nu}dx^{\mu}dx^{\nu},\qquad \mbox{with}\qquad
g_{\mu\nu}=\eta_{\mu\nu}+h_{\mu\nu},
\end{equation}
where $\eta_{\mu\nu}=\rm{diag}(1,-1,-1,-1)$ is the Minkowski first fundamental form, then its modification $h_{\mu\nu}$ is given by
\begin{eqnarray}\label{GR_lineal}
\Box {h}_{\mu\nu}
= -16\pi G \left({T}_{\mu\nu} - \frac{1}{2} \eta_{\mu\nu}\, T \right),
\end{eqnarray}
where
{$\Box\equiv \partial_{t^2}^2-\Delta$} denotes the d'Alembertian and  $T$ is the trace of the stress-tensor
\begin{equation}
{T}_{\mu\nu} = (\rho + p)\, u_{\mu}  u_{\nu} - p\,\eta_{\mu\nu},
\end{equation}
 defined in Minkowski spacetime, that is, 
\begin{eqnarray}
u_{\mu}=\frac{dx_{\mu}}{ds_{\rm Min}},\qquad \mbox{with}\qquad ds_{\rm Min}^2=\eta_{\mu\nu}dx^{\mu}dx^{\nu}.
\end{eqnarray}

On the other hand, taking the trace of Eq.~\eqref{GR_lineal} yields $\Box h = 16\pi G T$, allowing the field equation to be rewritten as
\begin{equation}\label{linear_equation}
\Box\!\left({h}_{\mu\nu}-\frac{1}{2}\eta_{\mu\nu} h\right)
= -16\pi G\,{T}_{\mu\nu}.
\end{equation}

Imposing the conservation law
$
\partial_{\mu}{T}^{\mu\nu}=0
$
in Minkowski spacetime and {taking the divergence of Eq.~\eqref{linear_equation} gives}
\begin{equation}
\Box \left[
\partial_{\mu}\left({h}^{\mu\nu}
-\frac{1}{2}\eta^{\mu\nu}h\right)
\right]=0.
\end{equation}

{With the asymptotic boundary conditions adopted throughout the construction, in particular the normalization of the gravitational potential to vanish at spatial infinity and the requirement of an asymptotically Minkowskian metric, the relevant homogeneous solution is set to zero. Hence,}
\begin{equation}
\partial_{\mu}\left({h}^{\mu\nu}
-\frac{1}{2}\eta^{\mu\nu}h\right)=0,
\end{equation}
which is precisely the linearized harmonic gauge condition.

Conversely, in General Relativity formulated in harmonic gauge, taking the divergence of the linearized field equations {directly yields
$$
\partial_{\mu}T^{\mu\nu}=0.
$$

Thus, in the weak-field regime, energy-momentum conservation and the harmonic gauge condition are mutually consistent, with the former determining the homogeneous equation for the gauge quantity and the chosen asymptotic conditions selecting the vanishing solution.}

\medskip

In summary, 
applying the spinozian concepts of \textit{conatus} and \textit{potentia} to nature, and 
starting from Galilean free fall together with the  Lorentz Transformation, we 
arrive---without postulating spacetime geometry---at an invariant structure that coincides exactly with linearised General Relativity in harmonic gauge. As emphasized by Weinberg~\cite{Weinberg1972}:
\begin{quote}
\textit{All the classical tests of General Relativity—the precession of planetary orbits, the deflection of light, and the gravitational redshift—can be derived from the linearized field equations.}
\end{quote}

{

\subsection{Description of the method}
Following Chapter II of \cite{Haro2026a}, we take as our starting point the line element derived for a strictly uniform gravitational field by enforcing free-fall kinematics:
\begin{equation}
    ds^2=(1+2gz)\,dt^2-(1+2gz)^{-1}\,d\mathbf{r}\cdot d\mathbf{r}.
\end{equation}

To describe an arbitrary static gravitational field, the most natural physical generalization consists in replacing the uniform term $gz$ with a general position-dependent Newtonian potential {$\Phi^{\rm stat}(\mathbf{r})$}. Rather than postulating a new spacetime metric ad hoc from abstract geometric principles, we adopt as a physically grounded working hypothesis the following conformastatic invariant \index{Conformastatic invariant} \cite{Synge1960}:
{\begin{equation}\label{conformastat}
    ds^2 = (1+2\Phi^{\rm stat}(\mathbf{r}))\,dt^2 - (1+2\Phi^{\rm stat}(\mathbf{r}))^{-1} d\mathbf{r}\cdot d\mathbf{r}.
\end{equation}}

{

\begin{remark}
The use of a restricted metric ansatz in the present reconstruction is deliberate and follows a methodological strategy with a clear precedent in Einstein's early development of General Relativity. In his 1912 investigations of the static gravitational field \cite{Einstein1912}, Einstein first determined the physical laws appropriate to a uniformly accelerated reference frame, obtained the corresponding line element $ds^2$, and, on the basis of the equivalence principle, transferred this structure to a uniform gravitational field, subsequently assuming that the same local form held for a static field. 

The present construction follows an analogous operational strategy: it begins in a physically controlled situation, determines the corresponding spacetime structure, and generalizes it to more complex gravitational settings. 
The methodological analogy should not be taken to imply that the two constructions are physically identical: what changes here is the foundational starting point, which is taken to be the dynamical law of free fall formulated in terms of the particle's proper time. 
The metric ansatz is therefore not intended as an arbitrary postulate of the final gravitational geometry, but as an intermediate operational structure whose symmetries (such as isotropy or plane/spherical invariance)
are dictated by the physical setup and whose functional form is constrained by the adopted dynamical principle.
Within this ansatz and the stated normalization conditions, the resulting invariant is subsequently tested against the weak-field equations of General Relativity.
\end{remark}

}

We next evaluate how moving, non-static, and pressureless matter distributions alter the background Minkowski invariant $ds^2$ within the weak-field regime.

In this weak-field approximation, we systematically restrict our analysis to terms linear in the gravitational potential. The justification for this truncation is fundamentally physical: to extend the static line element to moving sources, Lorentz transformations are applied to the individual elementary contributions associated with the material sources. However, a Lorentz boost is a purely kinematical operation defined on a flat Minkowski background; its scope is strictly confined to describing the motion of localized material sources.

The difficulty does not lie in the Lorentz transformation itself, which remains well defined beyond the linear regime, but in the linear superposition of independently boosted elementary contributions. At higher orders, nonlinear combinations of the perturbations encode the self-interaction of the gravitational field, so the elementary contributions can no longer be combined by simple linear superposition.
Consequently, the Lorentz transformation is applied exclusively to the linearized static invariant, thereby isolating the gravitational field generated solely by the motion of matter.

Thus, in a frame, namely $\Sigma$, where the masses are at rest, in the weak-limit,  we have
{\begin{eqnarray}\label{conformastat_weak}
  ds^2=(1+2\Phi^{\rm stat}({\bf r}))\,dt^2-(1-2\Phi^{\rm stat}({\bf r})) d{\bf r}\cdot d{\bf r},
\end{eqnarray}}
where the potential {$\Phi^{\rm stat}$} satisfies the Poisson equation {$\Delta\Phi^{\rm stat}=4\pi G\rho$.}
 
Let us consider another frame $\Sigma'$ moving with three-dimensional velocity $-{\bf v}'$ with respect to frame $\Sigma$. 
Equivalently, $\Sigma$ moves with velocity 
\begin{eqnarray}
   {\bf v}'=\frac{d{\bf r}'}{dt'}
\end{eqnarray}
with respect to $\Sigma'$.

Let $\gamma$ be the Lorentz factor
\begin{eqnarray}
   \gamma=\frac{1}{\sqrt{1-|{\bf v}'|^2}} .
\end{eqnarray}

Then, the Lorentz transformation that leaves the Minkowski interval invariant,
i.e., $ds_{\rm Min}^2=dt^2-d{\bf r}\cdot d{\bf r}=dt'^2-d{\bf r}'\cdot d{\bf r}'$, reads
{\begin{eqnarray}
  {\bf r}=\gamma\left[
    \frac{1}{\gamma}{\bf r}'+\frac{\gamma}{1+\gamma}({\bf v}'\cdot {\bf r}'){\bf v}'
    -{\bf v}'t'
    \right],\qquad 
    t=\gamma(t'-{\bf v}'\cdot {\bf r}').
\end{eqnarray}}
The key point is that, given two constants $a$ and $b$, the quadratic form
$a\,dt^2 + b\, d{\bf r}\cdot d{\bf r}$ transforms into
\begin{equation}
\gamma^2(a+b|{\bf v}'|^2)\, dt'^2
-2\gamma^2(a+b)({\bf v}'\cdot d{\bf r}')\, dt'
+ b\, d{\bf r}'\cdot d{\bf r}'
+ \gamma^2(a+b)({\bf v}'\cdot d{\bf r}')^2.
\end{equation}

Then, the invariant in $\Sigma'$, generated by a source moving with velocity ${\bf v}'$, can be written as
\begin{eqnarray}\label{LORENTZ}
ds^2 &=&
(1 - 2{\Phi}' + 4\Upsilon')\, dt'^2
- 8\, {\mathbf{N}'} \cdot d\mathbf{r}'\, dt'
- (1 - 2{\Phi}')\, d\mathbf{r}' \cdot d\mathbf{r}'
+ 4\, u'_{ij}dx'^idx'^j,
\end{eqnarray}
where the new potentials are functions of the   potential {${\Phi}'(t',{\bf r}')\equiv\Phi^{\rm stat}({\bf r}(t',{\bf r}') )$, which must satisfy Abraham's equation \index{Abraham's equation}\cite{Abraham1911}
\begin{eqnarray}
\Box {\Phi}' = -4\pi G {\rho}' \quad \mbox{with}\quad 
{\rho}'(t',{\bf r}')=\rho({\bf r}(t',{\bf r}').
\end{eqnarray}

The Lorentz-transformed static solution satisfies the corresponding wave equation. For an eternally uniformly moving source, this solution coincides with both the retarded and advanced representations; the distinction between these boundary conditions becomes relevant only for genuinely time-dependent source histories.

Therefore, 
for an eternally uniformly moving source, we can  identify the Lorentz-boosted potential $\Phi'(t',{\bf r}')$ with its retarded representation and write
\begin{eqnarray}
{\Phi}^{\rm ret}(t',{\bf r}') = -G\int \frac{{\rho}'(t'_{\rm r},\bar{\bf r})}{|{\bf r}'-\bar{\bf r}|}\, d\bar{V},
\end{eqnarray}}
where $t'_{\rm r} = t' - |{\bf r}'-\bar{\bf r}|$ is the retarded time.
The velocity ${\bf v}'$ enters through
\begin{eqnarray}
    \Upsilon' = \frac{{\Phi}^{\rm ret}}{1-|{\bf v}'|^2}, \qquad
    {\bf N}' = \frac{{\Phi}^{\rm ret}
    }{1-|{\bf v}'|^2} {\bf v}', \qquad
    {u}'_{ij} = \frac{{\Phi}^{\rm ret}
    }{1-|{\bf v}'|^2} v'_i v'_j.
\end{eqnarray}

These potentials satisfy the equation:
\begin{eqnarray}
\Box \Upsilon' = -\frac{4\pi G {\rho}'}{1-|{\bf v}'|^2}, \quad
\Box {\bf N}' = -\frac{4\pi G {\rho}'}{1-|{\bf v}'|^2} {\bf v}', \quad
\Box {u}'_{ij} = -\frac{4\pi G {\rho}'}{1-|{\bf v}'|^2} { v}'_i  {v}'_j.
\end{eqnarray}

At this point we remove the primes, and thus,  the invariant takes the form:
\begin{eqnarray}\label{LORENTZ1}
ds^2 &=&
(1 - 2{\Phi}^{\rm ret} + 4\Upsilon)\, dt^2
- 8\, {\mathbf{N}} \cdot d\mathbf{r}\, dt
- (1 - 2{\Phi}^{\rm ret})\, d\mathbf{r} \cdot d\mathbf{r}
+ 4\, u_{ij}dx^idx^j,
\end{eqnarray}

To verify the consistency of this results with conventional linearized General Relativity,
we consider the linearized equations \eqref{GR_lineal}. 
Setting $\mu=\nu=0$ in \eqref{GR_lineal} yields
\begin{eqnarray}
\Box h_{00}
=
8\pi G \rho
-
16\pi G \frac{\rho}{1-|{\bf v}|^2},
\end{eqnarray}
which is satisfied for $h_{00}=-2\Phi^{\rm ret}+4\Upsilon$.

For $\mu=0$ and $\nu=j$, since the covariant spatial component of the four-velocity under the $(+,-,-,-)$ signature is $u_j = - \gamma v_j$, the energy-momentum tensor component is $T_{0j} = -\rho \frac{v_j}{1-|\mathbf{v}|^2}$. Thus, the field equation takes the form
\begin{eqnarray}
\Box h_{0j}
=
16\pi G
\frac{v_j}{1-|{\bf v}|^2}\rho,
\end{eqnarray}
which, compared to the definition of ${\bf N}$, implies $h_{0j} = -4N_j$. This correctly reproduces the cross-term $2h_{0j} dt dx^j = -8 {\bf N} \cdot d{\bf r} dt$ present in the invariant~\eqref{LORENTZ1}.

Finally, for $\mu=i$ and $\nu=j$,
\begin{eqnarray}
\Box h_{ij}
=
-8\pi G \delta_{ij}\rho
-
16\pi G
\frac{v_i v_j}{1-|{\bf v}|^2}\rho,
\end{eqnarray}
which leads to $h_{ij}=2\Phi^{\rm ret}\delta_{ij}+4{u}_{ij}$.

Therefore, expression~\eqref{LORENTZ1} coincides with the result obtained in linearized General Relativity in the harmonic gauge,
when the stress tensor is considered in Minkowski space\footnote{As clearly explained in Misner, Thorne, and Wheeler's book \cite{MTW1973} — the linearised Einstein equations are obtained by linearizing the Ricci tensor while keeping the 
stress–energy tensor defined in Minkowski spacetime. This procedure is independent of the velocities of the sources.}, and thus, 
 the four-velocity is given by
\begin{eqnarray}
{\bf u}=\frac{1}{\sqrt{1-|{\bf v}|^2}}(1, {\bf v}).
\end{eqnarray}

{Next we construct the invariant for $N$
material objects with different masses and different constant velocities.

Note that the invariant (\ref{LORENTZ1}) has the structure 
$ds^2=ds^2_{\rm Min}+ds^2_{\rm perturbed}$, with 
\begin{eqnarray}\label{perturbed0}
  ds^2_{\rm perturbed}=
(- 2{\Phi}^{\rm ret} + 4\Upsilon)\, dt^2
- 8\, {\mathbf{N}} \cdot d\mathbf{r}\, dt
+ 2{\Phi}^{\rm ret}\, d\mathbf{r} \cdot d\mathbf{r}
+ 4\, u_{ij}dx^idx^j,
\end{eqnarray}
which in co-moving coordinates has the form (see eq. (\ref{conformastat_weak}))
\begin{eqnarray}\label{perturbed}
ds^2_{\rm perturbed}    =2\Phi^{\rm stat}({\bf r}_{\rm c})(dt_{\rm c}^2+ d{\bf r}_{\rm c}\cdot d{\bf r}_{\rm c}) \quad \mbox{with}\quad \Delta\Phi^{\rm stat}=4\pi G\rho,
\end{eqnarray}
where the subindex "${\rm c}$" denotes the comoving frame.

In the case of \(N\) material objects, the perturbed part of the line element can therefore be decomposed into elementary contributions, each of which is naturally written in the inertial coordinates in which the corresponding source is at rest:
\begin{eqnarray}
ds^2_{\rm perturbed}
=
\sum_{k=1}^N
2\Phi_k^{\rm stat}({\bf r}_k)
\left(
dt_k^2+d{\bf r}_k\cdot d{\bf r}_k
\right),
\end{eqnarray}
where \(k=1,\ldots,N\) labels the material sources and

$$
\Delta\Phi_k^{\rm stat}=4\pi G\rho_k.
$$

Each term represents the same physical contribution to the line element independently of the coordinates used to describe it. We may therefore express each elementary contribution in the common coordinates of the observer by applying the Lorentz transformation associated with the velocity of the corresponding source.

More precisely, to obtain the perturbed invariant in the coordinates of the observer $(t,\mathbf r)$, we apply to each elementary contribution the Lorentz boost associated with the velocity of the corresponding source. For each eternally uniformly moving source, the boosted static solution coincides with its retarded representation. We therefore obtain (\ref{perturbed0}) with
\begin{eqnarray}
    \Phi^{\rm ret}(t,{\bf r})=\sum_{k=1}^N\Phi_k^{\rm ret}(t, {\bf r})
    \quad \mbox{with}\quad
    \Phi_k^{\rm ret}(t,{\bf r})\equiv\Phi_k^{\rm stat}({\bf r}_k(t,{\bf r}))
    \Longrightarrow
    \Box
    \Phi_k^{\rm ret}(t,{\bf r})
    =-4\pi G\rho_k({\bf r}_k(t,{\bf r})),
\end{eqnarray}
and the same for the other potentials.

Therefore, for a family of different masses moving with different velocities, the invariant coincides with the one of General Relativity in the harmonic gauge (\ref{GR_lineal}), when the stress-energy tensor is given by
\begin{eqnarray}
    T^{\mu\nu}=\sum_{k=1}^N \rho_k u_k^{\mu}u_k^{\nu}.
\end{eqnarray}

In the case of $N$ infinitesimal masses, taking the limit $N\rightarrow\infty$ we obtain the case of a continuous distribution.

The construction can then be extended in two complementary ways. For fluids with pressure, one may directly use the linearized field equations with the general stress-energy tensor. Alternatively, within the operational construction developed here, one may consider non-constant velocities evaluated at the retarded event of each source element
 using  at the comoving frame
the following invariant}
\begin{equation}
ds^2 = (1 + 2\Phi_1)\,dt^2 - (1 - 2\Phi_2)\,d{\bf r}\cdot d{\bf r},
\end{equation}
where the scalar potentials satisfy
\begin{equation}
\Delta \Phi_1 = 4\pi G\,(\rho + 3p),
\qquad
\Delta \Phi_2 = 4\pi G\,[2(\rho + \,p)-(\rho+3p)],
\end{equation}
being $\rho+3p$ the \textit{active gravitational mass density} and $\rho+p$ the \textit{effective inertial mass density}
\cite{Haro2026,Haro2026a}.

\begin{remark}
For pressureless dust, energy-momentum conservation $\partial_{\mu}T^{\mu\nu}=0$ in Minkowski spacetime {implies that the individual fluid elements follow inertial trajectories. Their velocities are therefore constant along their worldlines, so that the present and retarded velocities coincide. The distinction becomes relevant} in the presence of pressure: internal stress gradients accelerate the fluid elements, {and the velocity entering} the Lorentz transformation must then be evaluated at the retarded event associated with each source element.
\end{remark}

This agreement demonstrates that, within the weak-field regime, the present formulation recovers the physical predictions of linearized General Relativity in the harmonic gauge while dispensing with an explicit geometric description in terms of spacetime curvature. In this approach, the geometric properties traditionally ascribed to spacetime emerge instead as an effective representation of the underlying dynamical interplay between matter and inertia.

Furthermore, in the static case, the invariant (\ref{conformastat}) automatically satisfies the harmonic gauge condition. Because this condition is preserved under Lorentz transformations, its boosted counterpart—describing moving sources—inherently obeys the same gauge constraint. This provides a natural explanation for why the expressions derived here for moving masses coincide exactly with those obtained within linearized General Relativity formulated in harmonic coordinates.

By consistently integrating the combined contributions of all moving sources into a unified line element, this procedure yields the same physical outcomes of harmonic-gauge General Relativity without ever deploying the full geometric machinery of the theory. Echoing Fock’s {viewpoint} \cite{Fock1964}, 
{from the perspective adopted in this essay, harmonic coordinates are especially useful because they provide the coordinate representation in which the dynamical structure of the invariant is most transparent. This should be understood as a physically motivated interpretative preference within the present framework, rather than as a claim that harmonic coordinates are uniquely preferred by General Relativity.}
Ultimately, this supports the view that, at least in the weak-field limit, the essential physical content of General Relativity can be fully captured by an appropriately defined field-theoretic invariant, consistent with both Galilean free fall and Lorentz invariance, without recourse to the full 
differential-geometric machinery of General Relativity.

}

\section{Gravitation for strong fields}
The framework developed in the preceding section provides a robust description of weak gravitational fields—faithfully capturing gravitational waves and relativistic source dynamics across all velocity regimes—from a foundational theoretical standpoint. 
{However, a purely linear description is incomplete from the standpoint of universal energy-momentum coupling.
At linear order, the gravitational field is sourced by the matter energy-momentum tensor on the Minkowski background, while the energy carried by the gravitational field itself contributes only at higher order.

}

Insisting on strict linearity explicitly breaks this universal coupling, {which would imply that matter generates gravity while gravitational energy does not.}
 To restore dynamical self-consistency, the field equations must be 
 {extended to include}
  this gravitational self-coupling, which inherently introduces non-linearity into the theory. In this section, the precise mathematical structure of this non-linear completion, which governs the transition from weak to strong fields without relying on arbitrary geometric postulates, will be determined.

In harmonic coordinates, the 
non-linear generalization of the linear equation \eqref{GR_lineal} takes the form
\begin{equation}\label{generalisation}
g^{\alpha\beta}\partial^2_{\alpha\beta}
{g}_{\mu\nu} +{U}_{\mu\nu}
=
-16\pi G \left({T}_{\mu\nu} - \frac{1}{2} {g}_{\mu\nu} T \right),
\end{equation}
where   
${T}_{\mu\nu}$ is the full matter energy-momentum tensor 
and ${U}_{\mu\nu}$ is a non--covariant term depending on ${g}_{\mu\nu}$ and its first derivatives. 

Remarkably, in harmonic coordinates, the Ricci tensor takes the form 
{\begin{eqnarray}
    R_{\mu\nu}=&&-\frac{1}{2}
g^{\alpha\beta}\partial^2_{\alpha\beta}
    g_{\mu\nu}
+\bar{R}_{\mu\nu} \end{eqnarray}}
where 
$\bar{R}_{\mu\nu}$ denotes the non-covariant expression built from ${g}_{\mu\nu}$ and its first derivatives.

When the invariant $ds^2$ represents a small perturbation of the Minkowski background, that is, ${g}_{\mu\nu}=\eta_{\mu\nu}+{h}_{\mu\nu}$, the linear approximation reduces to
\begin{equation}\label{Ricci1}
R_{\mu\nu} = -\frac{1}{2}\Box {h}_{\mu\nu},
\end{equation}
and {has the same principal differential structure as the linearized field equation (\ref{GR_lineal}).

 This motivates seeking a covariant non-linear completion of the linearized equations}.
{Under the assumptions 
---namely locality, dependence on the metric and its first two derivatives, second–order field equations, and compatibility with the covariant conservation of the energy–momentum tensor in four spacetime dimensions---
the Lovelock result selects the Einstein tensor, namely $G_{\mu\nu}=R_{\mu\nu}-\frac{1}{2}g_{\mu\nu}R$ where $R$ is the Ricci scalar,
up to a cosmological term, as the unique divergence-free symmetric rank-two tensor satisfying these requirements.
In the present framework, Minkowski spacetime is taken as the source-free reference configuration; this excludes a nonzero cosmological constant, so we set \(\Lambda=0\).
Therefore, taking into account the linearized form of the Ricci tensor (\ref{Ricci1}), the natural generalization of the linearized equation (\ref{linear_equation}) is
 \begin{eqnarray}
     G_{\mu\nu}=8\pi G T_{\mu\nu}.
 \end{eqnarray}

Finally, trace-reversing the field equations gives}
\begin{equation}\label{X}
R_{\mu\nu}=
8\pi G \left({T}_{\mu\nu}-\frac{1}{2}{g}_{\mu\nu}T\right),
\end{equation}
which are the equations presented by Einstein in November 1915
\cite{Einstein1916}.

In this formulation, the {Einstein} tensor is not introduced as an a priori geometric postulate; rather, it emerges as the unique covariant completion of the reduced wave operator governing the gravitational field, once energy–momentum conservation and second-order dynamical consistency are imposed.

After some algebra, the field equations can be written as
\begin{eqnarray}\label{main-equation}
{\Box_{g}}
{g}_{\mu\nu}
&=& -16\pi G\left(
({T}_{\mu\nu}+\bar{{t}}_{\mu\nu})
-\frac{1}{2}{g}_{\mu\nu} (T+\bar{t})
\right),
\end{eqnarray}
where {we have denoted by $\Box_{g}$ the Laplace-Beltrami operator $\frac{1}{\sqrt{-g}}\partial_{\alpha}\left(\sqrt{-g}g^{\alpha\beta}\partial_{\beta}\right)$, which in harmonic coordinates takes the form $g^{\alpha\beta}\partial^2_{\alpha\beta}$,
and}
$\bar{t}$ is the trace of the pseudo-tensor
\begin{equation}
\bar{t}_{\mu\nu}
\equiv
-\frac{1}{8\pi G}
\left(\bar{R}_{\mu\nu}-\frac{1}{2}{g}_{\mu\nu}\,\bar{R}\right),
\qquad \mbox{which} \qquad \bar{R}=\bar{R}_{\mu}^{\mu},
\end{equation}
which, {following the historical motivation provided by the first version of Einstein's field equations presented in 1915 \cite{Einstein1916}, we interpret \(\bar t_{\mu\nu}\), within this harmonic-coordinate formulation, as an effective energy-momentum pseudo-tensor for the gravitational field.
As usual for gravitational energy-momentum pseudo-tensors, \(\bar t_{\mu\nu}\) is coordinate-dependent and should not be regarded as a covariant local stress-energy tensor of the gravitational field.
 Its role here is instead to collect the nonlinear gravitational terms that appear as an effective source when Einstein's equations are written in harmonic coordinates as a wave equation. This interpretation is therefore tied to the chosen coordinate formulation rather than being a coordinate-independent local definition of gravitational energy-momentum.

}

{On the other hand, 
this pseudo-tensor should not be confused with the Landau--Lifshitz energy-momentum complex \cite{Landau},
namely $t_{\rm LL}^{\mu\nu}$,
which provides a different coordinate-dependent representation of the gravitational energy-momentum balance
\begin{eqnarray}\label{LL-conservation}
    \partial_{\mu}[(-g)(T^{\mu\nu}+t_{\rm LL}^{\mu\nu})]=0.
\end{eqnarray}

In fact, both pseudo-tensors are related 
as follows:
\begin{eqnarray}
    \bar{t}^{\mu\nu}=t_{\rm LL}^{\mu\nu}-\frac{1}{16\pi G}\left[
\left(g^{\mu\sigma}g^{\nu\gamma}-\frac{1}{2}g^{\mu\nu}g^{\sigma\gamma}\right)\Box_{g}g_{\sigma\gamma}
    +\frac{1}{(-g)}
\partial_{\sigma\gamma}^2H^{\mu\sigma\nu\gamma}\right],
\end{eqnarray}
where 
$H^{\mu\sigma\nu\gamma}=(-g)(
g^{\mu\nu}g^{\sigma\gamma}
-g^{\mu\gamma}g^{\sigma\nu}
)$ is a tensor used to write the field equations as \cite{Poisson}
\begin{eqnarray}
\partial_{\sigma\gamma}^2H^{\mu\sigma\nu\gamma}=16\pi G (-g)(T^{\mu\nu}+t_{\rm LL}^{\mu\nu}),
\end{eqnarray}
and has the remarkable property
\begin{eqnarray}
\partial_{\sigma\gamma\mu}^3H^{\mu\sigma\nu\gamma}=0,
\end{eqnarray}
which leads to the conservation law 
(\ref{LL-conservation}).
}

This expression {\color{red} (\ref{main-equation})} makes explicit that Einstein's equation can be interpreted as a non-linear wave equation \index{Non-linear wave equation} in which the gravitational field not only responds to matter but also acts as its own source. 
{In this formulation, the nonlinear terms represented by \(\bar t_{\mu\nu}\) make explicit the self-coupling of the gravitational field.

}

\begin{remark}
Expressing Einstein's field equations in harmonic coordinates illuminates their intrinsic dynamical mechanics, transcending the conventional geometric paradigm. This framework aligns directly with the field-theoretic approach to gravitation pioneered by Feynman, Gupta, Thirring, and Deser, wherein General Relativity arises as the consistent non-linear completion of a massless spin-2 field universally coupled to total energy-momentum—including its own. Within this perspective, Einstein's equations represent the covariant architecture that enforces  dynamical self-consistency through the field's necessary non-linear self-couplings. 
{In fact, the field equations in the form of a wave equation (\ref{main-equation}) has the same structure as the field equation used by Thorne and Kovacs in \cite{Thorne} (see for instance eq. (10b)), and its solution can be written as a integral equation via the Green function, namely $G_{g}$, in a curved spacetime, as follows
\begin{eqnarray}
    g_{\mu\nu}({\bf x})=\int G_{g}({\bf x};{\bf x'})\tau_{\mu\nu}({\bf x'})d^4{\bf x'},
\end{eqnarray}
where we have used the notation
\begin{eqnarray}
    \tau_{\mu\nu}\equiv -16\pi G\left(
({T}_{\mu\nu}+\bar{{t}}_{\mu\nu})
-\frac{1}{2}{g}_{\mu\nu} (T+\bar{t})
\right).
\end{eqnarray}

}

Consequently, spacetime geometry ceases to be an \textup{a priori} foundational postulate. It emerges instead as the effective manifestation of a self-sourcing field dynamics.
\end{remark}

The findings above prompt a fundamental reassessment of a long-standing conceptual paradigm: the tendency to view gravitation as an autonomous geometric architecture imposed upon an otherwise passive spacetime. The analysis presented here offers a compelling alternative. Rather than positing gravity as an \textit{a priori} geometric agency, it emerges naturally from a remarkably minimal set of foundational physical principles: the universality of Galilean free fall and the requirement that physical trajectories, in the absence of non-gravitational forces, {optimize} proper time. From this strictly dynamical standpoint, Einstein's field equations are neither axiomatic postulates nor geometric {assumptions;
under the conditions stated above, 
 they arise, 
  {within the stated framework}, as the self-consistent nonlinear completion selected by the Lovelock assumptions.}
Geometry, in this view, forfeits its role as the ontological substrate of the theory, serving instead as its descriptive language—a mathematical encoding of how physical clocks and rods respond to gravitational interactions.
Within this framework, {harmonic coordinates acquire a more than merely technical role: they provide the coordinate setting in which the underlying dynamical interpretation developed here becomes most transparent.

}

It is precisely within these coordinates that the weak-field regime {reveals its character} as the direct relativistic extension of Newtonian gravity, {without requiring} auxiliary hypotheses or interpretative {constructions beyond those developed in the present framework}. In this light, General Relativity recovers its conceptual {foundations}: not as a doctrine governing the curvature of an abstract manifold, but as a concrete dynamical theory of time, inertia, and free fall. Dynamics and geometry converge in this way into a unified and self-consistent picture, in which the mathematical formalism directly manifests---rather than obscures---its physical and epistemological foundations.

\subsection{Heuristic Machian Relationalism and Gravitational Self-Energy}
Inspired by Mach's critique of absolute space, gravitation is framed here as a relational phenomenon connecting matter, fields, and the modulation of inertia. Within this framework, the fundamental invariant $ds$ is not an autonomous self-subsistent geometric entity; rather, it operates as a physical descriptor encoding how clocks and trajectories are dynamically affected by the distribution of mass and energy. However, unlike radical Machian doctrines—which demand that spacetime and inertia vanish entirely in the absence of matter—our field-theoretic approach treats Minkowski spacetime as the unperturbed, reference baseline. In the complete absence of gravitational sources, the invariant naturally reverts to its flat limit ($ds^2_{\rm Min} = \eta_{\mu\nu} dx^\mu dx^\nu$), which represents the unmodulated state of proper time and free fall rather than a Machian total void.

This perspective closely mirrors Riemann's original methodological vision, in which the metric properties of space were not treated as primitive, geometrically given, but as structures whose physical origin lies outside geometry. Operationally, once the invariance of the speed of light is postulated, proper time assumes {operacional} primacy, while spatial distances are derivative constructs, which are reconstructed from light-signal exchanges, and are no more primitive geometric notions \cite{Landau}.

However, the non-linear structure of General Relativity introduces a subtle ontological dilemma when evaluating the so-called ``vacuum solutions''. In conventional relativistic physics, a vacuum is technically defined by the local vanishing of the matter stress-energy tensor ($T_{\mu\nu}=0$). Yet, when Einstein's field equations are cast in harmonic coordinates, their underlying field-theoretic architecture comes into sharp focus:
\begin{equation}
g^{\alpha\beta}\partial^2_{\alpha\beta}{g}_{\mu\nu} = -16\pi G\left( \bar{{t}}_{\mu\nu} -\frac{1}{2}{g}_{\mu\nu} \bar{t} \right).
\end{equation}

This representation makes explicit that the condition \(T_{\mu\nu}=0\) does not {imply that the gravitational field dynamics become trivial. Within the harmonic-coordinate formulation adopted here, the nonlinear gravitational terms remain on the right-hand side through the effective pseudo-tensor \(\bar t_{\mu\nu}\).}

A standard objection may be that $\bar{t}_{\mu\nu}$ is not a fully covariant tensor, its non-zero character is coordinate-dependent. Coordinate dependence is undeniable, but this property holds strictly within the harmonic gauge. As argued throughout the present essay, the harmonic gauge is not merely a mathematical convenience issue, but a naturally preferred, inertial-like framework where the equations explicitly reveal their field-theoretic structure. In this formulation,
the pseudo-tensor {provides an effective representation of the nonlinear gravitational contributions to the source term.}

This distinction carries crucial epistemological consequences. Vacuum solutions---such as source-free gravitational waves, the exterior Schwarzschild metric, or anisotropic vacuum cosmologies like the Kasner and Bianchi models (without matter)---are not dynamical fluctuations arising from the ``geometry of a literal nothingness''. Instead, they either anchor their curvature to localized singular sources (such as $r=0$ in Schwarzschild), import energy through non-trivial asymptotic boundary conditions at infinity ($\mathscr{I}^-$), or unfold from initial temporal boundaries ($t=0$). In all these cases, where material stress-energy is locally absent ($T_{\mu\nu}=0$), 
{these configurations are represented by nonlinear gravitational contributions that enter the effective source through $\bar t_{\mu\nu}$.}
Within a strictly relational ontology, a true physical vacuum would require the total mathematical annihilation of the entire right-hand side. Consequently, flat Minkowski spacetime emerges merely as a trivial relational limit---a formal baseline that carries no autonomous physical existence outside of pure abstraction. The fact that General Relativity allows the field to act as its own source through $\bar{t}_{\mu\nu}$---even when $T_{\mu\nu}=0$---suggests that the standard geometric interpretation conceals an equivalent field-theoretic reality. From this viewpoint, orthodox GR attributes to geometry what, in the harmonic field formulation, appears explicitly as the self-energy of the gravitational field.

The perspective championed here invites one to resolve this tension by stripping vacuum solutions of any claim to geometric substantivism. Rather than demonstrating that a literal geometrical void possesses autonomous dynamics, these solutions show that a relativistic spin-2 field, upon achieving its necessary non-linear completion, acquires the capacity to channel its own energy-momentum in the absence of material boundaries. Ultimately, the persistence of $\bar{t}_{\mu\nu}$ as an active source in a formal vacuum strongly indicates that General Relativity can be more naturally interpreted not as the curvature of an abstract void, but as the dynamics of a self-interacting physical field driven by energy.

This realization prompts a critical re-evaluation of the orthodox consensus. John Archibald Wheeler  encapsulated the standard geometric interpretation with his much celebrated sentence \cite{MTW1973}:
\begin{quote}
\textit{Space tells matter how to move; matter tells space how to curve.}
\end{quote}

Despite its extraordinary pedagogical power, this formulation fosters a substantivist reading in which spacetime appears as an independent entity acting upon matter. Yet, as demonstrated by Einstein's hole argument \cite{Stachel2014}, active hole diffeomorphisms---transformations that rearrange manifold points within source-free regions while leaving material sources untouched---generate no new physical states. This clearly reveals that the differentiable manifold, taken in isolation, carries no direct physical content; 
{physical observables are instead tied to point coincidences and field relations.} Consequently, attributing autonomous physical meaning to source-free geometries---or imposing non-trivial global topologies on a material void (such as in Taub-NUT-type constructions, where a gravitomagnetic monopole charge acts as an effective topological source exciting the field)---constitutes a form of topological substantivism. It reifies the points of the bare manifold, directly violating the relational lesson of the hole argument by treating the mathematical map as an ontological container.

Viewed through the lens of the harmonic field equations, Wheeler's aphorism appears as conceptually incomplete. By absorbing the field's non-linear self-interaction into the geometric left-hand side of Einstein's equations, the standard formalism creates the illusion of an empty, passive spacetime reacting to matter. Once the active dynamical role of $\bar{t}_{\mu\nu}$ is recognized, a more accurate field-theoretic conclusion emerges:

\begin{quote}
\textit{Matter and the field’s own energy govern the unfolding of gravity,  {while gravity,  in turn, shapes the structure of proper time,  which}  sculpts the inertial paths of matter.}
\end{quote}

From this perspective, geometry forfeits its foundational ontological status, serving instead as the formal mathematical language through which the coupled dynamics of matter and gravitational self-energy are articulated.

\section{Conclusions}
The development of General Relativity introduced a profound tension between physical insight and mathematical formalism---a tension that {can remain obscured by the standard geometric formulation.}

This inclination to construct physical theories axiomatically, {in a manner reminiscent of the axiomatization of pure mathematics}, was strongly championed by David Hilbert. Indeed, his 1900 Sixth Problem explicitly called for the axiomatization of the physical sciences, a program he later extended to gravitation in his 1915--1917 investigations \cite{Hilbert1916,Branding}. Einstein {likewise developed Special Relativity} from two fundamental postulates: the Principle of Relativity and the constancy of the speed of light. {In General Relativity, the Equivalence Principle played a similarly foundational role, while the subsequent development of the theory led to the adoption of the differential-geometrical framework required to formulate the gravitational field equations.}

Although such axiomatic and geometric formulations are operationally sound and mathematically efficient, they risk obscuring the physical insight from which the theory originally sprang. This tension is central to the present reconstruction, which recovers the operational and dynamical content underlying the geometric formalism---a perspective echoing Einstein's own frequent warnings regarding the limitations of purely formal mathematical reasoning.

To make this operational grounding explicit, one must re-examine the concept of inertia itself. In the standard geometric framework, inertial motion is subordinated to spacetime curvature, reduced to a mere mathematical consequence of parallel transport along a geodesic. In contrast, the present reconstruction restores the physical origin of inertia through the \textit{conatus} of a body to
{locally} maximize its experienced proper time. Straight-line motion  ceases to be an abstract primitive postulate of the Cartesian and Newtonian tradition; it naturally emerges as the zero-gravity limit
 of a more fundamental operational principle of free fall.

However, once the advanced tools of absolute differential calculus became indispensable to formalize the field equations, the conceptual focus progressively shifted from operational physics toward geometry. Over time, this geometric interpretation grew dominant, encouraging a substantialist view of spacetime as an autonomous physical container. The analysis presented here demonstrates that an alternative operational route is fully feasible, showing how the familiar geometric formalism naturally emerges from underlying dynamical principles at the final stage of the derivation.

By {questioning} the idea of an autonomous curved spacetime container, {the present reconstruction} restores the primacy of the territory over the map. The hole argument, together with the dynamical equivalence exhibited by teleparallel formulations of gravity \cite{Weitzenbock1923}, {provides} historical and theoretical {motivation for questioning} whether geometric structures should be assigned autonomous ontological status. From a historical perspective, this construction suggests that an operational route to the weak-field structure of General Relativity was already available within the conceptual framework of Special Relativity, once the appropriate static invariant {had been} identified. In this sense, the present work outlines a path that was historically available, although never explicitly pursued.

In this framework, gravitation is {reconsidered not as a primitive geometric structure, but from its physical and operational foundations}. It is {interpreted} as a physical manifestation of Spinoza’s \textit{potentia}---the capacity of matter to modulate the local rate of proper time. {By} contrast, the motion of freely falling particles is governed by their \textit{conatus}---the intrinsic endeavor to {optimize} experienced proper time along a worldline. Dynamics and temporality are thus unified at the operational level: gravity does not act {by bending a pre-existing} spatial substance; rather, {it manifests itself through a modulation of clock rates, while freely falling particles follow extremal worldlines.}

This shift from a substantive conception of geometry to operational dynamics carries profound epistemological implications. Echoing Newton's \textit{hypotheses non fingo}, the framework deliberately refrains from assigning primitive ontological status to geometric structures. Curvature and torsion are interpreted as descriptive tools that encode relations between physical processes rather than primitive physical entities. Likewise, the metric tensor and its invariants are not tangible physical entities but relational constructs governing motion and measurement. This perspective aligns with the lessons of the hole argument and offers a clean, flat-background starting point where linearized field equations in harmonic coordinates emerge directly from the kinetic state of the sources.

Ultimately, this work presents a fundamental reconsideration of the nature of spacetime. Geometry is not the starting physical postulate but the final mathematical synthesis emerging from operational first principles. The resulting formulation is mathematically equivalent to the standard theory within its domain of validity, yet it offers a distinct conceptual narrative: proper time, inertia, and free fall precede geometry, which appears only when the requirements of general covariance, energy-momentum conservation, and internal dynamical consistency demand a non-linear completion of the weak-field equations. 
It is at this point that the {Einstein} tensor emerges as the unique {divergence-free covariant structure 
within the
class of theories
satisfying the assumptions of Lovelock’s theorem, up to a cosmological term.
With the Minkowski reference structure adopted here, the cosmological term is excluded, leaving Einstein's field equations as the resulting non-linear completion.}

Geometry is therefore not abandoned, {but reinterpreted}: General Relativity remains, {in full}, a geometric theory, but geometry is its final destination {rather than} its genesis.

\begin{acknowledgments}
JdH  {is supported by the Spanish grant PID2025-168310NB-I00 funded by MICIU/AEI/10.13039/501100011033 and by FSE+.}
  EE is partly supported by the program Unidad de Excelencia María de Maeztu, CEX2020-001058-M, and by AGAUR, project 2021-SGR-00171.
\end{acknowledgments}

\end{document}